\documentclass[12pt]{iopartamsmath}

\usepackage[english]{babel}
\usepackage{siunitx}
\AtBeginDocument{\RenewCommandCopy\qty\SI}
\usepackage{caption, subcaption}
\usepackage[printonlyreused]{acronym}
\usepackage{booktabs}
\usepackage{multirow}
\usepackage[square, numbers]{natbib}

\usepackage{amsmath}    
\usepackage{graphicx}
\usepackage[colorlinks=true, allcolors=blue]{hyperref}
\usepackage{bm}
\usepackage{lipsum}
\usepackage{physics}
\usepackage{ulem}

\DeclareSIUnit\EDMunits{\SI{}{\it{e}~\centi\metre}}
\DeclareUnicodeCharacter{4E00}{一}
\DeclareUnicodeCharacter{59DC}{姜}
\DeclareUnicodeCharacter{541B}{君}

\usepackage{color}

\begin{document}

\title{Low-noise environment for probing fundamental symmetries }

\author{F. J. Collings, N. J. Fitch\footnote{Present address: Infleqtion, 3030 Sterling Cir. Boulder CO 80301, USA}, R. A. Jenkins, J. M. Dyne, E. Wursten, M. T. Ziemba, X. S. Zheng, F. Castellini, J. Lim, B. E. Sauer, M. R. Tarbutt\footnote{E-mail: m.tarbutt@imperial.ac.uk}\footnote{Corresponding author}}
\address{Centre for Cold Matter, Imperial College London, London, SW7 2AZ, United Kingdom}

\begin{abstract}
We present the design and characterization of a low-noise environment for measuring the electron's electric dipole moment (EDM) with a beam of molecules. To minimize magnetic Johnson noise from metals, the design features ceramic electric field plates housed in a glass vacuum chamber. To suppress external magnetic noise the apparatus is enclosed within a cylindrical four-layer mu-metal shield with a shielding factor exceeding $10^6$ in one radial direction and $10^5$ in the other. Finite element modelling shows that the difference between these shielding factors is due to imperfect joints between sections of mu-metal. Using atomic magnetometers to monitor the magnetic field inside the shield, we measure noise below 40~fT/$\sqrt{{\rm Hz}}$ at 1~Hz and above, rising to 500~fT/$\sqrt{{\rm Hz}}$ at 0.1~Hz. Analytical and numerical studies show that residual magnetic Johnson noise contributes approximately 13~fT/$\sqrt{{\rm Hz}}$. The background magnetic field averaged along the beamline is maintained below 3~pT, with typical gradients of a few nT/m. An electric field of 20~kV/cm is applied without discharges and with leakage currents below 1~nA. Each magnetometer measures the magnetic field correlated with the direction of the applied electric field with a precision of 0.11~fT in 104 hours of data. These results demonstrate that the apparatus is suitable for measuring the electron EDM with precision at the $10^{-31}$~e~cm level. The design principles and characterization techniques presented here are broadly applicable to precision measurements probing fundamental symmetries in molecules, atoms, and neutrons.

\end{abstract}

\section{Introduction}

Experiments with neutrons, atoms and molecules can probe new symmetry-violating physics with extraordinary precision~\cite{DeMille2017, Safronova2018}. Experiments with molecules are especially notable. They are being used to measure parity (P) violating interactions in the nucleus~\cite{Altuntas2018}, electric dipole moments that violate both parity and time-reversal symmetry (T)~\cite{Hudson2011, Andreev2018Short, Roussy2023}, P,T-violating nuclear Schiff moments~\cite{Grasdijk2021} and magnetic quadrupole moments~\cite{Hutzler2020, Denis2020, Ho2023}. The electron's electric dipole moment (EDM, $d_e$) is tiny in the Standard Model~\cite{Ema2022} but many orders of magnitude larger in most extensions of the Standard Model, including supersymmetric extensions, due to the ubiquity of new T-violating interactions in these theories~\cite{Engel2013, Chupp2019}. Measurements of the electron EDM using molecular beams and trapped molecular ions have set the upper bound $|d_e| < 4.1 \times 10^{-30}~e$~cm at the 90\% confidence level, and this result already constrains broad classes of new physics at energies of order 10~TeV~\cite{Andreev2018Short, Roussy2023}.
New techniques being developed promise a new generation of advanced experiments with even greater physics reach~\cite{DeMille2024}. Molecules can now be cooled to ultracold temperatures by direct laser cooling~\cite{Fitch2021b}, and electron EDM experiments using laser-cooled diatomic and polyatomic molecules are being developed~\cite{Alauze2021, Fitch2020b, Anderegg2023}. Ultracold molecules can also be assembled from ultracold atoms, including radioactive species with exceptional sensitivity to nuclear P,T-violating physics~\cite{Fleig2021,Klos2022}. Measurements with polar molecules isolated in cryogenically-grown inert ices also promise spectacular sensitivity~\cite{Li2023b}.

Atoms and molecules with an unpaired electron and no nuclear spin, such as the $^{174}$YbF isotopologue used in this work, are mainly sensitive to two T-violating effects -- the electron EDM, $d_e$, and a T-odd electron-nucleon interaction characterized by the parameter $C_s$. When the nucleus has a spin, the experiments may also be sensitive to the nuclear Schiff moment, magnetic quadrupole moment or nuclear-spin-dependent T-violating electron-nucleon interactions. These experiments typically search for a symmetry-violating energy level splitting induced by an applied electric field, analogous to the symmetry-conserving Zeeman splitting induced by a magnetic field. The measurements are difficult because they may require large electric fields and because the tiny energy difference of interest has to be distinguished from a Zeeman splitting that may be orders of magnitude larger. Some systems provide a natural immunity to these difficulties. For example, some polar diatomic molecules in $^{3} \Delta_1$ states are highly sensitive to the electron EDM, can be fully polarized using small electric fields, and have magnetic $g$-factors close to zero~\cite{DeMille2001,Vutha2010}. Some polyatomic molecules are also easily polarized and states can be found that are sensitive to the electron EDM but relatively insensitive to magnetic fields~\cite{Kozyryev2017b, Takahashi2023, Anderegg2023}. Other systems do not have all these properties but instead have other advantages, e.g. they may be easy to cool and trap, or may have very high sensitivity to the new physics of interest, or especially long coherence times. These experiments require exceptional control over electric and magnetic fields. It is often necessary to apply high electric fields while ensuring low leakage currents. Magnetic fields usually need to be controlled and measured carefully because excess magnetic noise may ruin the experimental sensitivity and changes in magnetic fields that correlate with electric field reversals are a source of systematic error. Such experiments often need to be enclosed in multi-layer magnetic shields~\cite{Yashchuk2013} or in magnetically-shielded rooms~\cite{Altarev2015Short}, and ultra-precise magnetometry may be necessary throughout a large volume, e.g.~\cite{Abel2020}.

In this paper, we describe a low-noise environment developed for measuring the electron EDM using a beam of laser-cooled YbF molecules~\cite{Fitch2020b}. This species is promising because it has high intrinsic sensitivity to $d_e$~\cite{Kozlov1997}, is laser-coolable~\cite{Lim2018,Alauze2021}, and has been used previously to make EDM measurements~\cite{Hudson2002,Hudson2011}. It also has some drawbacks: it is vulnerable to noise and systematic errors generated by magnetic fields, and the experiment requires a high electric field. Consequently, we have designed the experiment to reduce external magnetic noise to a suitable level, minimize magnetic Johnson noise, apply a large electric field, and maintain ultra-high vacuum. To satisfy these requirements, the apparatus features ceramic electric field plates housed in a glass vacuum chamber, surrounded by atomic magnetometers and a four-layer magnetic shield. The design goals include an electric field of 20~kV/cm, magnetic noise below 100~fT/$\sqrt{\rm{Hz}}$ at 0.5~Hz and above, and the capability of measuring electric-field-correlated magnetic fields with a precision of 0.1~fT in a few days of measurement time. We present the design, characterize its performance, and determine the EDM sensitivity that could be reached in this apparatus. We expect many aspects of our design to be useful to other experiments including other electron EDM measurements using cooled or trapped molecules~\cite{Aggarwal2018Short, Anderegg2023}, those using radioactive species~\cite{Simsarian1996, Arrowsmith-Kron2024Short}, measurements using atoms or molecules to probe symmetry-violating effects in the nucleus~\cite{Grasdijk2021, Parker2015Short, Zheng2022, Yang2023, Allmendinger2019}, and experiments with neutrons~\cite{Schmidt-Wellenburg2017}.

\section{Overview}

\subsection{Overview of experiment}

The work described here is part of a project to measure the electron EDM ($d_e$) using ultracold YbF. The experiment uses molecules in the ground electronic, vibrational and rotational state which consists of a pair of hyperfine components with angular momenta $F=0$ and $F=1$. It is convenient to introduce the notation $\ket{0} = \ket{F=0,m_F=0}$, $\ket{\pm 1} = \ket{F=1,m_F=\pm 1}$, $\ket{+} = \frac{1}{\sqrt{2}}\left(\ket{+1} + \ket{-1} \right) $ and $\ket{-} = \frac{i}{\sqrt{2}}\left(\ket{+1} - \ket{-1} \right) $. A cryogenic buffer gas source produces a pulsed beam of YbF molecules with a mean speed of about 150~m/s~\cite{Truppe2017c, White2024}. Laser cooling is applied in the two transverse directions to collimate the beam, so that it can pass through the rest of the apparatus with very little divergence~\cite{Lim2018, Alauze2021}. The cooled molecules are prepared in $\ket{0}$ and then pass into the `interaction region' which is the topic of this paper. Here, they are subjected to an electric field $\vec{E}$ and magnetic field $\vec{B}$, both applied in the same direction ($z$), and nominally constant. Using stimulated Raman adiabatic passage (STIRAP), the molecules are transferred to $\ket{+}$, which then evolves in time $\tau$ to $\frac{1}{\sqrt{2}}\left(e^{-i\phi}\ket{+1} + e^{i\phi}\ket{-1} \right) = \cos\phi\ket{+} - \sin\phi\ket{-}$, where 
\begin{equation}
\phi =\frac{1}{\hbar} \int_{0}^{\tau} (g \mu_{\rm B} B_z - d_e E_{\rm eff})\,dt.
\label{eq:phi}
\end{equation}
Here, $g=1$ is the magnetic $g$-factor and $E_{\rm eff} = E_{\rm eff}^{\rm max} \eta(E)$ is an effective electric field. It is the product of a species-dependent factor which for YbF is $E_{\rm eff}^{\rm max} \approx -26$~GV/cm, and the polarization factor $\eta(E)$ which is a function of the applied electric field. At $E=20$~kV/cm, $\eta = 0.693$. In the presence of the strong electric field along $z$, the molecules have almost no sensitivity to magnetic fields perpendicular to $z$~\cite{Hudson2002}, so only $B_z$ appears in (\ref{eq:phi}). A second STIRAP transfers $\ket{+}$ back to $\ket{0}$, but does nothing to $\ket{-}$. After leaving the interaction region, we measure the populations in $F=0$ and $F=1$ which are proportional to $\cos^{2}\phi$ and $\sin^{2}\phi$, thus determining $\phi$. By reversing the relative directions of $B$ and $E$, we isolate the part of $\phi$ that is due to $d_e$. Fluctuations in $B_z$ on the timescale of this reversal cause fluctuations in $\phi$ that potentially mask the signal of interest, so these must be carefully controlled. Much of this paper relates to this task.

\subsection{Design overview}

\begin{figure}[p]
    \centering
    \includegraphics[width=\textwidth]{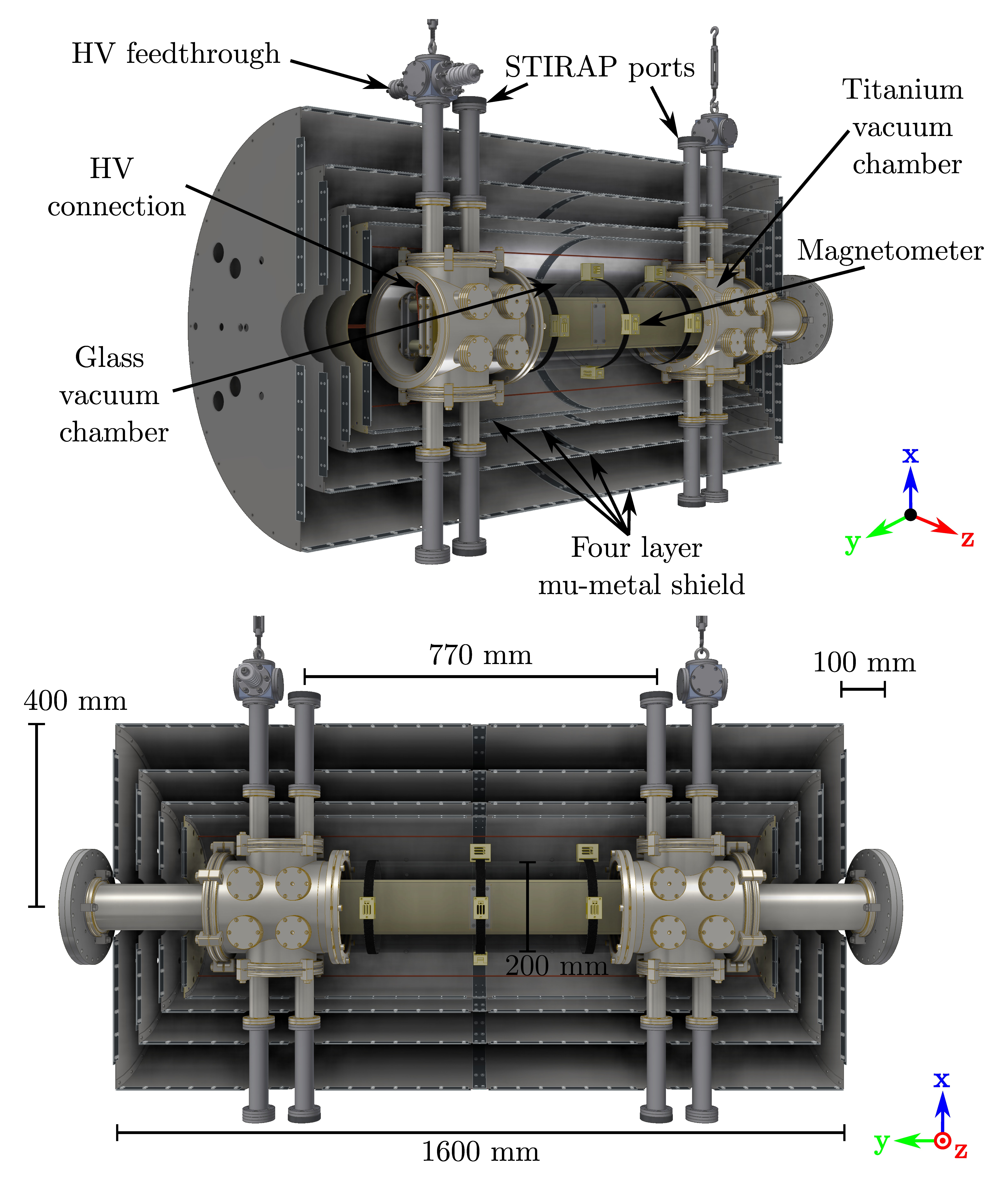}    \caption{Overview of the interaction region, showing the electric field plates, vacuum chambers, magnetometers and magnetic shields. Gravity is along $-x$. Molecules travel along $y$. Electric field is along $z$.}
    \label{fig:CADInteractionRegion}
\end{figure}

Figure \ref{fig:CADInteractionRegion} illustrates the design of the interaction region. Key drivers of the design are the shielding of laboratory magnetic noise and the control of magnetic Johnson noise arising from thermal fluctuations of electrons in conductors close to the molecular beam. At the centre of the apparatus are a pair of electric field plates made from aluminium oxide coated with titanium nitride. They are inside a vacuum tube of outer diameter 200~mm made from thermally pre-stressed borosilicate glass. This tube is sealed to titanium vacuum chambers at both ends using pre-baked\footnote{Baked under vacuum at approximately 110$^{\circ}$C for about 12 hours.} radial viton o-rings that fit tightly over the cylindrical surface of the glass and are pressed against custom retaining rings welded into the Ti chambers. The Ti chambers house the high voltage feedthroughs, the STIRAP laser access ports, and the structures that support the plates. The vacuum chambers have not been baked and the pressure is approximately $1\times10^{-7}$~mbar, presumably limited by outgassing and the relatively poor conductance to the turbomolecular pumps situated beyond the interaction region. No leaks were detected with a helium leak checker, sensitive at the $10^{-10}$~mbar~l/s level.

An array of zero-field optically pumped atomic magnetometers surrounds the glass vacuum tube in order to map the magnetic field in the interaction region. The whole interaction region is enclosed within a four-layer mu-metal magnetic shield. The shields are a set of nested cylinders, each constructed from four cylindrical sections connected together with mu-metal bands and terminated with end-caps. The vacuum chamber is supported from below at the two flanges immediately outside the shields, and from above via the ports that house the high voltage feedthroughs. To apply a uniform magnetic field throughout the interaction region, current-carrying wires run along the length, and internal to, the innermost layer of shielding. 

The interaction region is modular. The electric field plates and the cylindrical parts of the shields are fabricated in sections that are easily connected together. In figure~\ref{fig:CADInteractionRegion} there are two sections, but this can be extended to make a longer interaction region in the future. The glass tube can be manufactured to any length up to 10~m, and will always connect to the same two titanium chambers.

\section{Electric field}

\subsection{Construction}

\begin{figure}[p]
    \centering
    \includegraphics[height=\dimexpr\textheight-2\baselineskip-\parskip-.2em-\abovecaptionskip-\belowcaptionskip\relax]{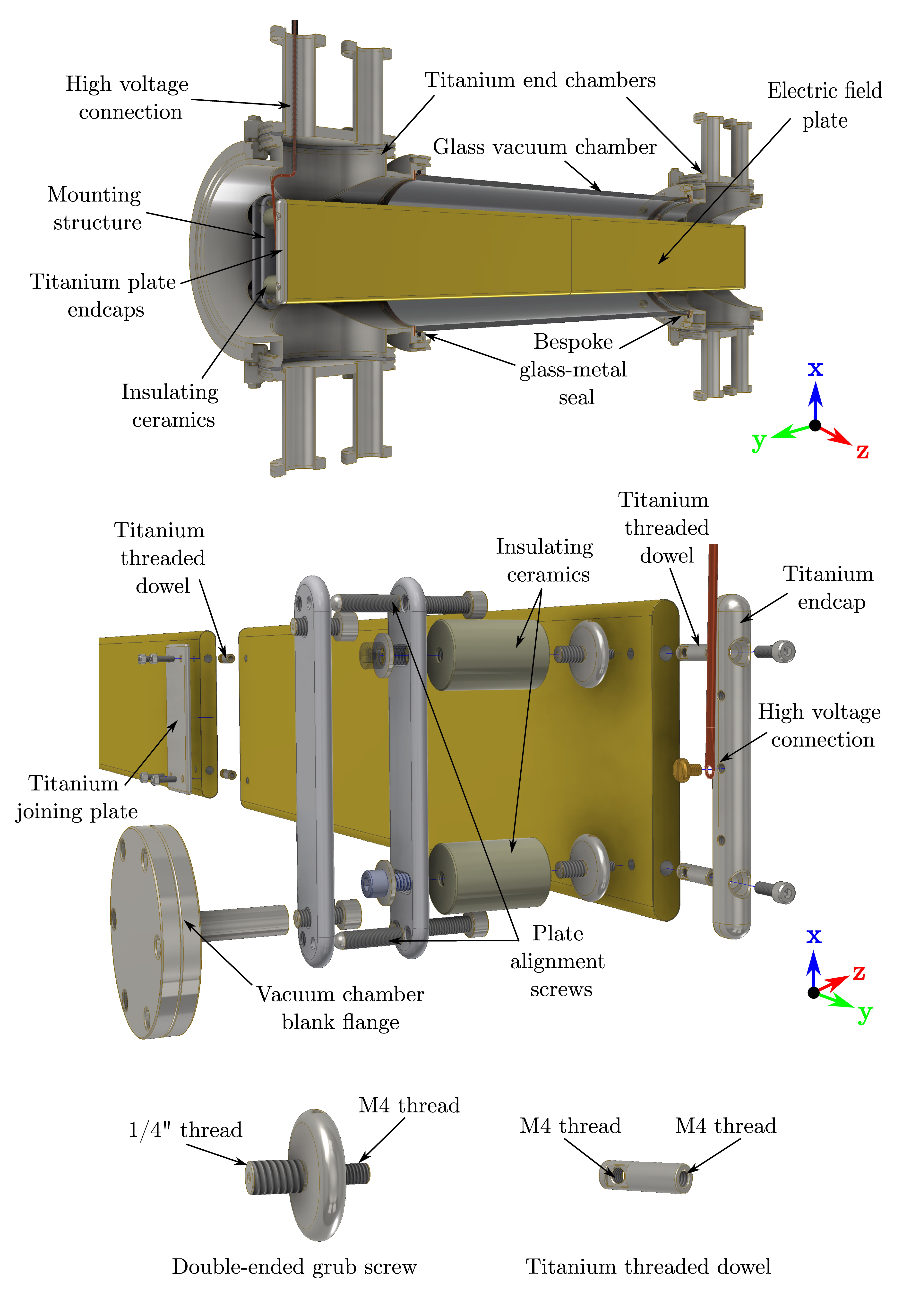}
    \caption{Construction of the electric field plates showing how two modules are connected together and how the plates are supported at the two ends.}
    \label{fig:CADPlates}
\end{figure}

Figure \ref{fig:CADPlates} shows the construction of the electric field plates. The plates need to be electrical conductors, so that the field can be applied, but the amount of conducting material should be small enough to keep the Johnson noise low. To this end, the plates are built from slabs of aluminium oxide (Al$_2$O$_3$), each 550~mm long, 120~mm high and 12~mm thick. The top and bottom edges are filleted with a radius of 5~mm to avoid sharp edges that could cause electric discharge. The slabs are coated with an interface layer of pure titanium, 50~nm thick, followed by a layer of titanium nitride (TiN), approximately 1~$\mu$m thick and applied using pulsed-DC reactive sputtering. TiN is a smooth, hard, chemically inert ceramic material with an electrical conductivity similar to that of titanium. The coating lowers the secondary electron emission coefficient of the bare aluminium oxide, helping to suppress electric discharge~\cite{Suharyanto2007}. The slabs have flat ends so that they can be connected together. Each slab has two 6.03~mm holes at the flat ends (parallel to $y$), intercepted by smaller blind holes from the back (parallel to $z$). The connection is made using titanium dowel pins, 5.95~mm in diameter, that are pushed into the holes in the flat ends of the slabs. A magnified view of a dowel is given in the figure. Each dowel has two orthogonal M4 threads: an axial thread at one end (parallel to $y$), and a thread machined radially (parallel to $z$) through the dowel close to the other end. Opposing dowels are connected together with an M4 titanium grub screw so that when the two slabs are pushed together the two radial threads in the dowels align with the holes at the back of the slabs. The connection is then secured using a titanium joining plate and four M4 titanium screws that screw into the radial threads. The joining plate ensures good mechanical and electrical connections. Multiple slabs can be joined together this way, though we currently only use two to produce plates with a total length of 1100~mm.

Once constructed to the desired length, the field plates are terminated at either end with rounded titanium endcaps to eliminate sharp corners, and to make the high voltage connections. These connect to the plates using the same dowels  described above and shown enlarged in the figure. Two steatite ceramic insulators, 25.4~mm in diameter and 38.1~mm long, with blind central threads, connect to the back of the plates with grub screws that screw into the cross-threaded dowels. Because the steatite insulators have imperial threads, whereas the dowels are metric, we use the titanium imperial-to-metric double-ended grub screws shown in the figure. The insulators then connect to the grounded support structure which in turn connects to the flanges of the titanium vacuum chamber. All components of the support structure are titanium. Once both electric field plates are installed, their precise positions can be controlled using fine threaded screws in the support structures. These are adjusted to make the plates parallel and to control the gap between them. We fixed the plate separation to be 18.24~mm, identical at all four corners to within the measurement uncertainty of 0.02~mm. To make the electrical connections, two high voltage feedthroughs enter via the same cube at the top of the machine (see figure~\ref{fig:CADInteractionRegion}). They connect to the titanium endcaps via 3~mm diameter copper rods running parallel to one another and threaded with ceramic beads. This design ensures that charging currents run antiparallel and in close proximity, which lowers the magnetic field they produce and reduces the likelihood that charging currents magnetize the layers of mu-metal shielding they pass through.

\subsection{Leakage currents}

\begin{figure}
    \centering
    \includegraphics[width=0.9\textwidth]{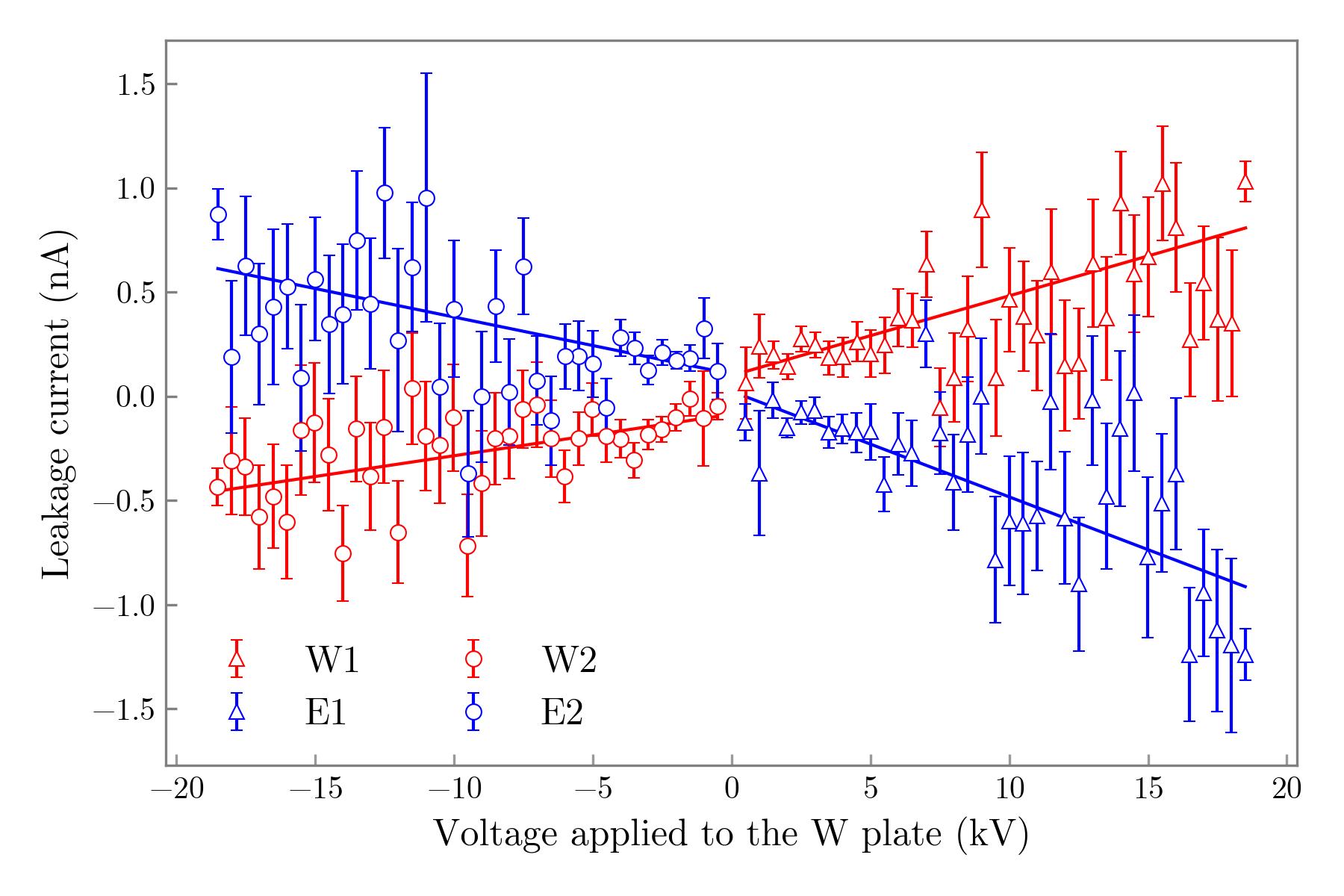}
    \caption{Leakage currents measured at each plate (labelled W and E) as a function of the applied voltage. The measurement is made for two polarities, labelled 1 when W is positive and E is negative, and 2 for the opposite case. The currents flowing to W and E are shown in red and blue, using triangles for polarity 1, and circles for polarity 2.}
    \label{fig:leakage}
\end{figure}

Electrical currents flowing to the plates, or between them, are of great concern because they can produce a magnetic field that reverses with the reversal of the electric field, mimicking an EDM. When the electric field is reversed, the acquisition of EDM data ceases until the charging currents have settled to a suitable level, and the timescale for this is chosen through the choice of series resistor between the high voltage power supplies and the plates. We use a 30~M$\Omega$ series resistor and measure a charging time constant of 20~ms. This gives good control over the charging currents whose influence can be reduced to a negligible level. More worrying are leakage currents that flow between the plates or from each plate to ground, typically leaking across the surfaces of high voltage insulators. We measure these currents using ammeters that float at high voltage and are connected to the high voltage feedthroughs (air side), in series between the power supplies and the electric field plates~\cite{Sauer2008}. They sense the current, convert it to a frequency, and transmit the data to a computer by optical fibre. 

We found that leakage currents increased over the course of several months and were correlated to humidity, sometimes even exceeding the 100 nA upper range of the ammeters on humid days. We attribute this to water vapour affecting the ceramic insulators of the high voltage feedthroughs, either forming a film on the surface or being absorbed into the body of the ceramic. To reduce this problem, we heat the feedthroughs to a constant temperature of 350~K. This has resulted in stable and reproducible results for the leakage currents. Figure \ref{fig:leakage} shows how the leakage currents, $i$, depend on the applied voltage, $V$. The plates are labelled West (W) and East (E) and are charged to $+V$ and $-V$ respectively. For these data, the polarity was reversed by swapping the cables from the two power supplies to the two plates (we call them polarities 1 and 2). The currents are consistent with a linear dependence on $V$ with gradients $G=i/V$ of $G_{W1}=(38\pm 5)$~pA/kV, $G_{E1}=(-51\pm 7)$~pA/kV, $G_{W2}=(20\pm 4)$~pA/kV and $G_{E2}=(-27\pm 5)$~pA/kV, where the subscripts indicate the plate and the polarity. The nominal operating voltage is $\pm 18.2$~kV, producing a field of 20~kV/cm. At this voltage, the average leakage current correlating with polarity is about 620~pA.

If these leakage currents are at the feedthroughs, which is likely, they are harmless, but we cannot rule out currents flowing along the plates. A full analysis of possible current paths is beyond the scope of this paper, but it is useful to consider some scenarios. First, we note that the two plates are connected to the feedthroughs at the same end, so if there are currents on both plates they must flow in opposite directions. Second, since the plates have a uniform surface, it seems most likely that currents would flow in roughly uniform sheets. We consider a uniform sheet of current flow in the $y$ direction along a single plate. This can only produce a $B_z$ at the molecules if they are offset from the centre of the plates in the $x$ direction. For example, an offset of 1~mm results in $B_z/i = 0.2$~fT/nA. Note that equal and opposite sheets of current flowing on the two plates produce no $B_z$, even when the molecules are offset from the centre. The current can be more dangerous if it follows a special, localized path. The edges of the plates potentially form special paths, so we hypothesize a current that flows along the top edge of one plate and back along the bottom edge of the other. This generates $B_z /i = 6.6$~fT/nA. Helpfully, the magnetometers will see this field too. In particular, one of the magnetometers is placed at exactly the same distance from the plate edge as the molecules are from the plate edge, so the magnetometry can potentially rule out an $E$-correlated $B_z$ arising from such a conspiratorial current path.

\subsection{Patch potentials}

Ideally, the electric field should point along $z$ throughout the machine. Spatial variations in the field direction can lead to the accumulation of a geometric phase, equal to the solid angle traced out by the electric field vector in the molecule's rest frame~\cite{Tarbutt2009}. This geometric phase may be a source of noise if it fluctuates, or a source of systematic error if it changes when the electric field is reversed. Variations of the electrostatic potential across the surface of the field plates -- known as patch potentials -- can give rise to changes in electric field direction and an associated geometric phase. When combined with a bend in the field plates a systematic error can arise~\cite{Kara2012}.

\begin{figure}[tb]
    \centering
    \includegraphics[width=0.75\textwidth]{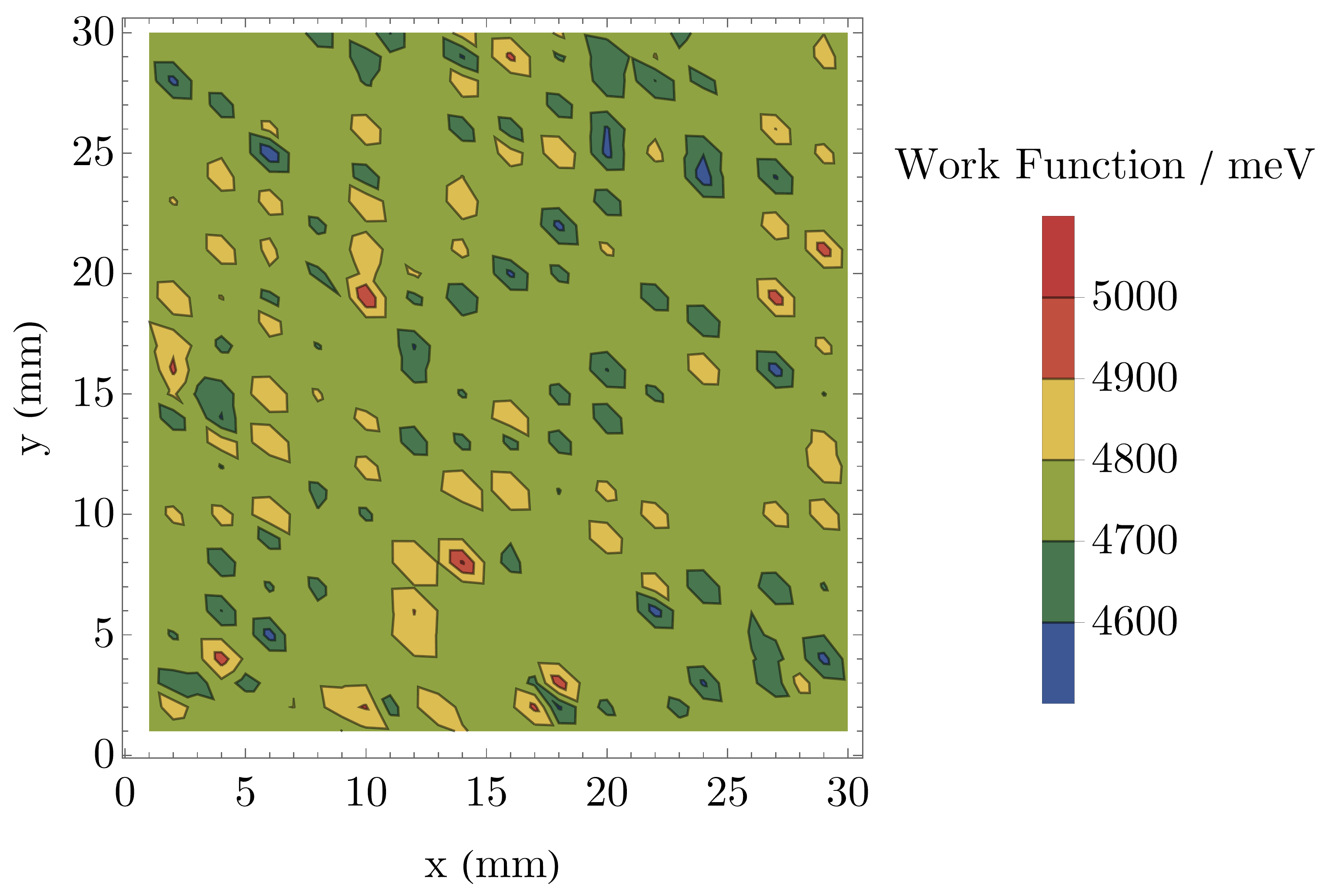}   
        \caption[]{A scan of the work function of a test sample of TiN-coated alumina. This contour plot shows a mean work function of \SI{4753}{\milli\eV} with an RMS variation across the sample of \SI{54}{\milli\eV}.}
    \label{fig:WorkFunctionPlateScan} 
\end{figure}

We prepared a test sample of TiN-coated Al$_2$O$_3$ and measured the variation of the work function using a Kelvin probe, calibrated using a pure silver sample as a reference. Figure~\ref{fig:WorkFunctionPlateScan} shows the variation in the work function across a 30~mm x 30~mm area. The mean value of the work function is $4753 \pm 10$~meV, where the uncertainty is the estimated accuracy of the probe. The root mean square (rms) variation of the work function is 54~meV. This is ten times the variation found for the silver sample. The patches seen in figure~\ref{fig:WorkFunctionPlateScan} have a characteristic size scale of 1~mm, too small to have a significant effect on the electric field at the molecules, which are several mm away from the plates. 

A previous analysis~\cite{Kara2012} considered aluminium field plates which were measured to bend by about $\pm 0.5$~mrad when mounted~\cite{Hudson2007}. It was supposed that the electric field first rotates by 1~mrad around $x$ due to the bend, then rotates around $y$ as molecules enter the region occupied by a 1~V  patch covering one quarter of the surface of one plate, followed by reverse rotations as the bend angle reverses and the molecules leave the region of the patch. For an applied field of 20~kV/cm and $\tau=20$~ms, this example produces a systematic error of $1.6 \times 10^{-32}$~e~cm. Our alumina plates are flat to within 20~$\mu$m and have a substantially higher elastic modulus than aluminium plates, so the bend angle is likely to be smaller. Moreover, the 1~V patch used for this estimate is far larger than anything we have measured. So we conclude that this is not a significant source of error.

\section{Magnetic field}

Magnetic noise can limit the sensitivity of the EDM measurement. The noise comes from several sources. The first is the laboratory magnetic noise, which is far too large and has to be reduced using several layers of magnetic shielding. Second is the noise on the applied magnetic field produced by a current-carrying coil, setting a constraint on the noise of the current supply. Third is magnetic Johnson noise due to fluctuating currents in nearby conductors. Finally, the magnetic shields themselves can be a source of noise due to thermal fluctuations of the magnetic domains. The molecules have almost no sensitivity to magnetic fields perpendicular to the applied electric field, so our main concern is fluctuations of $B_z$~\cite{Hudson2002}. 

In section \ref{Sec:ShieldDesign} we describe the design of the magnetic shields. Then, in section \ref{Sec:NoiseImpact} we study the impact of magnetic noise and calculate how large it can be if it is not to limit the precision of the experiment. Finally, in the remaining sections, we describe the methods used to control and measure the background magnetic field and its noise, and show that we are able to reach the required level.

\subsection{Shield design}\label{Sec:ShieldDesign}

\begin{figure}[tb]
    \centering
    \includegraphics[width=\textwidth]{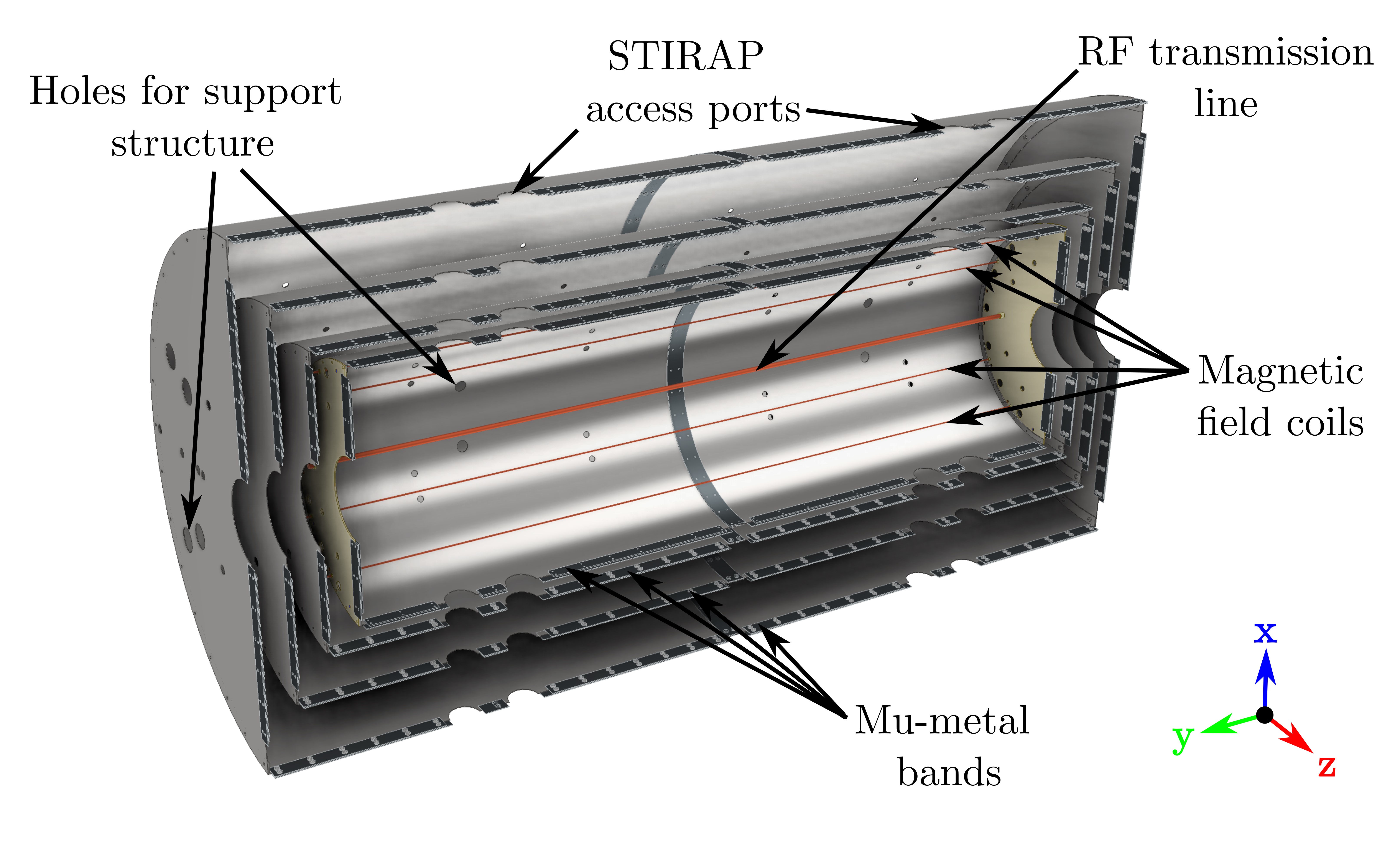}
    \caption{Design of the four-layer magnetic shield. A half section is shown.}
    \label{fig:CADShields}
\end{figure}

\begin{figure}[tb]
    \centering
    \includegraphics[width=\textwidth]{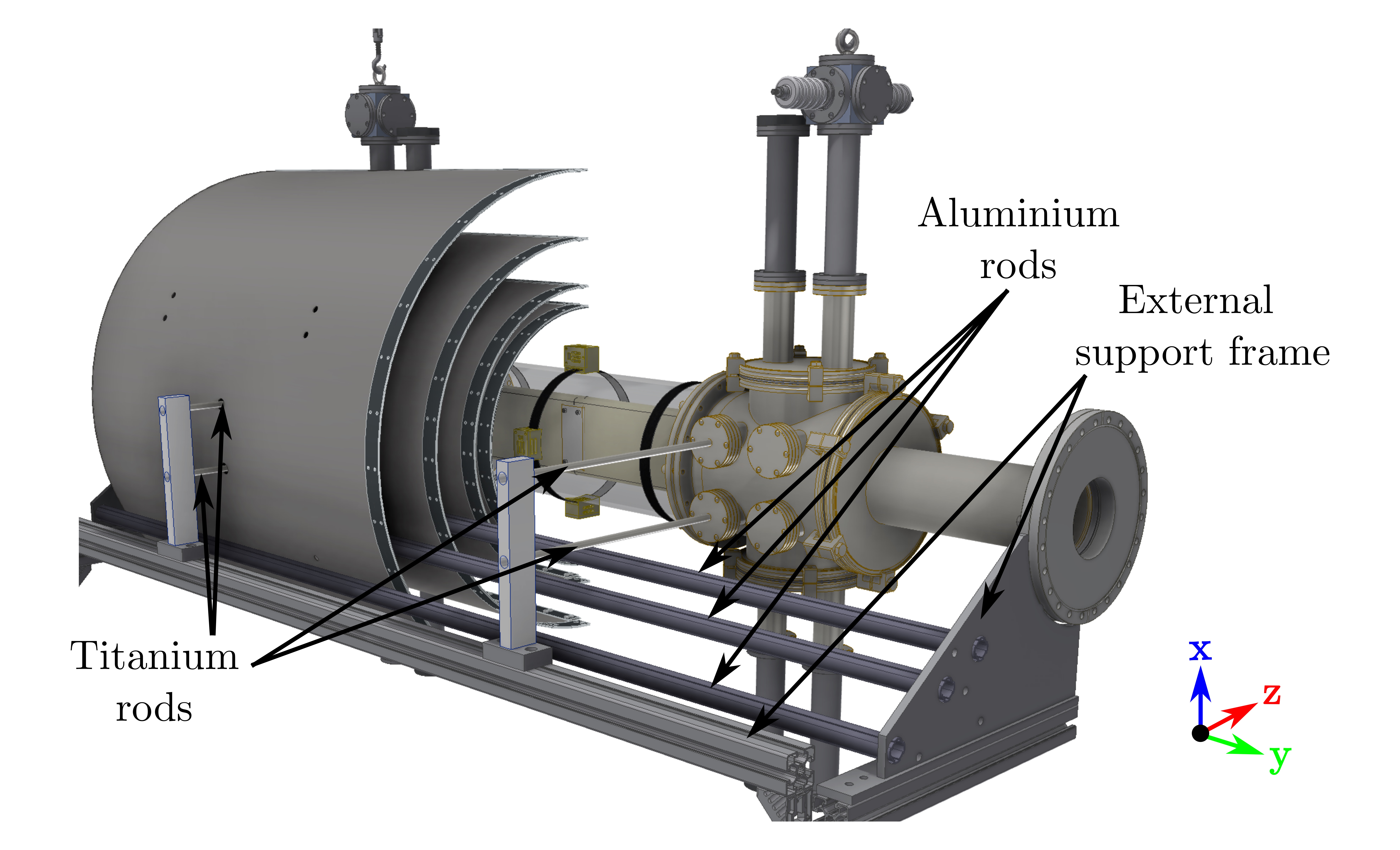}
    \caption{Mounting structure used to support the shields and vacuum chamber. Quarter sections of the shields are shown. The inner shield is supported by the titanium rods and the three outer shields by the aluminium rods. The chamber is supported from below at the large flanges outside the shields, and from above at the cube where the high voltage feedthroughs are mounted. An identical set of supports, not shown, are used on the other side of the apparatus (i.e. mirrored in the $xy$-plane).}
    \label{fig:ShieldMounting}
\end{figure}

Figure \ref{fig:CADShields} shows the design of the four-layer mu-metal shield. Once constructed, each shield is a closed cylinder. Their radii are 0.20, 0.23, 0.30 and 0.40~m, and their lengths are 1.3, 1.4, 1.5 and 1.6~m. Each cylinder is built from four curved sections and two flat endcaps. Where the curved sections meet, they are connected by joining bands. The bands that run along the length of the cylinder (parallel to $y$, see figure \ref{fig:CADShields}) are flat -- a mu-metal band with self-clinching nuts (sometimes called swage nuts or pem nuts) runs along the inside surface of the shield, ensuring the continuity of the mu-metal, while aluminium bands with clearance holes run along the outside to add rigidity. The innermost layer uses 2~mm thick mu-metal bands with tapped holes, instead of the pem nuts, to avoid the remanent magnetization that those fasteners could introduce.  To complete a cylinder, curved mu-metal bands with self-clinching nuts run around the circumference, along the inner surface (see figure \ref{fig:ShieldMounting}). The curved sections of the cylinders terminate with tabs where the endcaps connect directly. All connections are made with brass screws. 

At the ends of each cylinder there are holes, 120~mm in diameter, for the vacuum tube to pass through. In the curved surfaces, holes of 50~mm diameter provide access to the STIRAP ports and the high voltage feedthroughs. Figure \ref{fig:ShieldMounting} shows how the vacuum chamber and shields are supported. The innermost shield is supported by titanium rods that run parallel to $z$ and are attached to the CF40 vacuum flanges at one end and the external support frame at the other end. These rods pass through holes of 18~mm diameter in the shields (visible in figures \ref{fig:CADShields} and \ref{fig:ShieldMounting}). The other three shields rest on aluminium rods parallel to $y$ that pass through holes of 40~mm diameter in the endcaps of the two outer shields (see figure \ref{fig:CADShields}), and connect to the external support frame at both ends (see figure \ref{fig:ShieldMounting}). 

It is imperative to eliminate electrically conducting loops that enclose the shield. Such loops inductively couple oscillating magnetic fields into the shielded region. When we first built the shields, we accidentally had some of these ground loops, and they caused the shielding factor to decrease with increasing frequency, which is opposite to the expected behaviour of a shield. To eliminate these paths, we insulated all parts of the supporting structure of the shields. Specifically, all components passing through the holes in the shields were electrically insulated. Only then did we measure a shielding factor that increased with frequency.

To reduce the residual magnetic field inside the shield and improve the shielding performance, we apply a degaussing procedure, typically once per day. Degaussing coils run along the length of each shield, inside and outside, so that the material of the shield is enclosed. In each coil, we apply an ac current with a frequency of 5~Hz. The current ramps up to 3~A in 3~s, and is then reduced linearly over a period of 300~s. Once the current reaches zero a switch is opened, breaking the circuit so that oscillating magnetic fields are not inductively coupled into the shield. The degaussing procedure is applied to each layer of mu-metal in turn, starting with the outermost layer and moving inwards. After all layers have been degaussed, we leave the shield to rest for 20 minutes before making any measurements. The magnetometers show that the internal magnetic field relaxes on this timescale. 

To apply a magnetic field in the interaction region, eight wires run along the length of the innermost shield, lying on a circle of radius 180~mm. To minimize the amount of metal, while ensuring sufficient rigidity, the wires are copper tubes with an inner diameter of 3~mm and outer diameter of 4~mm. Four of the wires are oriented at $\pm 30^{\circ}$ to the $x$-axis and are used to apply a uniform field along $z$. The other four are oriented at $\pm 30^{\circ}$ to the $z$-axis and are used to apply a uniform field along $x$. They produce $5.0 \pm 0.1$~pT/$\mu$A. Finite element modelling suggests that the field is uniform to 1 part-per-million throughout the relevant volume. Two copper tubes of 8~mm diameter run parallel to the $y$-axis at $(x,z) = (0,\pm 135)$~mm. They serve as a radio-frequency transmission line for driving hyperfine transitions when the pulse of molecules is at any position along the length of the apparatus. This is a convenient way to build maps of the electric and magnetic fields along the beamline~\cite{Hudson2007}.

A set of 8 zero-field optically-pumped magnetometers\footnote{QZFM Gen-3 from Quspin} measures the magnetic field inside the inner shield. These magnetometers are attached to the outer surface of the glass vacuum tube at $r=\pm 115$~mm, and distributed along its length (they can be seen in figure \ref{fig:CADInteractionRegion}). Each measures the field along two orthogonal axes, and the set are configured to provide $\{5,3,8\}$ measurements of $\{B_x,B_y,B_z\}$. The magnetometer noise floor is 15~fT/$\sqrt{{\rm Hz}}$ above 3~Hz, but is not specified for the lower frequencies of most interest to us. The magnetic field outside the shields is measured using a 3-axis fluxgate magnetometer\footnote{Mag-03 from Bartington} with a range of $\pm 100~\mu$T and a noise floor of 6~pT/$\sqrt{{\rm Hz}}$ at 1~Hz.

\subsection{Impact of magnetic noise}\label{Sec:NoiseImpact}

We would like to know how fluctuations in the magnetic field impact the uncertainty in the measurement of $d_e$. Recall that the experiment measures the phase $\phi$ given by (\ref{eq:phi}). Let $\phi_{B}$ be the magnetic part of the phase, which is proportional to the magnetic field in the $z$-direction, integrated over the flight time of the molecules, $\tau$. The variance in this phase due to magnetic noise is~\cite{Munger2005}
\begin{equation}
    \sigma_{\phi_B}^2 = \left(\frac{\mu_{\rm B} \tau}{\hbar}\right)^2 \int_0^{\infty} B_{{\rm n},z}^2(f) \frac{\sin^2(\pi f \tau)}{(\pi f \tau)^2}d f,
    \label{eq:phase_variance_B}
\end{equation}
where $B_{{\rm n},z}(f)$ is the spectral density of the magnetic noise in the $z$-direction at frequency $f$, and we have set $g=1$. To make sure that magnetic noise does not compromise sensitivity, the variance given by (\ref{eq:phase_variance_B}) should be smaller than the variance due to quantum projection noise, which is $1/N_{\rm mol}$ for a phase measurement using $N_{\rm mol}$ molecules.  The ${\rm sinc}^2$ function in (\ref{eq:phase_variance_B}) rolls off the sensitivity to magnetic noise at frequencies above $1/\tau$. In the limit where the noise at lower frequencies is approximately constant, the integral evaluates to $B_{{\rm n},z}^2(0)/(2\tau)$ so we obtain
\begin{equation}
\sigma_{\phi_B} = \frac{\mu_{\rm B}B_{{\rm n},z}(0) \sqrt{\tau/2}}{\hbar}.
\label{eq:phase_deviation_B_low}
\end{equation}
In this limit, our magnetic noise requirement is
\begin{equation}
    B_{{\rm n},z}(0) < \frac{\hbar}{\mu_{\rm B}}\sqrt{\frac{2}{N_{\rm mol}\tau}}.
    \label{eq:B_noise_limit}
\end{equation}
For example, if $N_{\rm mol} = 10^6$ molecules per shot and $\tau = 2 \times 10^{-2}$~s, the requirement is $B_{{\rm n},z}(0) < 114$~fT/$\sqrt{{\rm Hz}}$.

This is not yet the complete picture because the electric field direction is modulated in the experiment, and the EDM is the part of the phase that correlates with this modulation, $\phi_E$. This method is sensitive to magnetic noise at the modulation frequency but not to noise at other frequencies. Suppose we make $N$ consecutive measurements of $\phi$  at equal time intervals. The state of the electric field switch for each measurement is defined by a balanced waveform $W_E$, which is a list of $N$ values, each either $+1$ or $-1$, with a mean of zero. The magnetic field during measurement $k$ is $B_0 + \beta_k$, where $B_0$ is a constant applied field and $\beta$ is the fluctuating part whose mean tends to zero for large $N$. We are interested in the uncertainty in $\phi_E$ arising from the magnetic noise $\beta$. It is~\cite{Hudson2014}
\begin{equation}
    \delta_{\phi_E} = \left(\frac{\mu_{\rm B} \tau}{\hbar}\right) \frac{1}{N} \sqrt{\sum_{k=1}^N \left|F[\beta]_k\right|^2 \left|F^{-1}[W_E]_k \right|^2},
    \label{eq:sigma_phiE_magnetic}
\end{equation}
where we have introduced the discrete Fourier Transform
\begin{equation}
    (F[x])_k = \frac{1}{\sqrt{N}} \sum^N_{m=1} x_m e^{-i\frac{2\pi m k}{N}}.
\end{equation}
Note that the sum in (\ref{eq:sigma_phiE_magnetic}) is proportional to $N$, so $\delta_{\phi_E}$ scales as $1/\sqrt{N}$ as expected for a standard error. We see from (\ref{eq:sigma_phiE_magnetic}) that each Fourier component of the magnetic noise contributes to the uncertainty in proportion to the Fourier component of the switching waveform at that frequency. In the case where the electric field is switched at a single frequency, $f_E$, (\ref{eq:sigma_phiE_magnetic}) gives the same result as (\ref{eq:phase_deviation_B_low}) but with $B_{{\rm n},z}(0)$ replaced with $B_{{\rm n},z}(f_E)$. Since the noise tends to fall off with increasing frequency, it is best to switch the electric field as rapidly as possible. In practice, since many other parameters are switched in the experiment, the switching waveform typically contains several Fourier components~\cite{Hudson2014}.

\subsection{Magnetic noise from conductors and shields --- analytical estimates}

Magnetic Johnson noise arises from the thermal agitation of charges in conductors. The power spectrum of the noise is independent of frequency at low frequencies $f$, and falls off as $1/f^2$ at high frequencies. The transition between the two regimes is at frequency $f_c = R/(2\pi L)$ where $L$ and $R$ are the inductance and resistance of the conductor surface. For most geometries, we find $f_c$ above 1~kHz, but we are interested in noise at much lower frequencies. Thus, we focus on the frequency-independent part. The magnetic noise at a distance $z$ above an infinite conducting slab of material of thickness $t$, resistivity $\rho$, and temperature $T$, has been considered in Ref.~\cite{Varpula1984}. At low frequencies, the magnetic noise in the $z$ direction (normal to the surface of the slab) is very well approximated as
\begin{equation}
    B_{{\rm n},z} = \mu_0 \sqrt{\frac{k_{\rm B} T t}{8\pi \rho z(t+z)}}.
    \label{Bnoise_slab}
\end{equation}
It is desirable to have non-magnetic conductors of high resistivity, so Ti and TiN are good choices, both having $\rho \approx 5 \times 10^{-7}$~$\Omega$\,m. Suppose we make a pair of electric field plates from one of these materials, with a gap of 10~mm between them, and evaluate the Johnson noise half way between the plates, remembering to multiply the above result by $\sqrt{2}$ since there are two plates. At room temperature, the noise will be $B_{{\rm n},z} = 370$~fT/$\sqrt{{\rm Hz}}$ when $t=10$~mm reducing to $B_{{\rm n},z} = 6.4$~fT/$\sqrt{{\rm Hz}}$ when $t=1$~$\mu$m. Comparing these numbers to the requirement found from (\ref{eq:B_noise_limit}) we see that self-supporting metal electrodes produce too much noise, whereas an insulating material coated with a thin layer of one of these materials makes a suitable electrode. This is the reason for choosing electric field plates made from Al$_2$O$_3$ coated with TiN. Similar considerations suggest that a titanium vacuum tube of diameter 200~mm produces too much noise. The noise is acceptable when the interaction region is housed inside a glass tube with titanium chambers at the two ends.

\begin{figure}[t]
    \centering
    \includegraphics[width=0.5\textwidth]{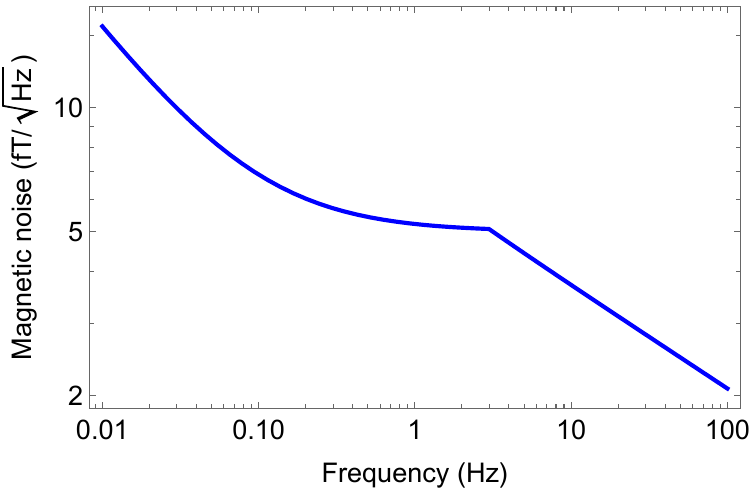}
    \caption[]{Noise spectrum due to the inner magnetic shield, calculated using Eqs.~(\ref{Bnoise_shield}) and (\ref{Bnoise_ratio}).  }
    \label{fig:ShieldNoiseAnalytical}
\end{figure}

The magnetic noise due to high permeability magnetic shields has been calculated by Lee and Romalis~\cite{Lee2008}. Using the fluctuation-dissipation theorem, they calculate the noise due to both current and magnetization fluctuations of the material. The noise due to current fluctuations (Johnson noise) on the axis of a cylindrical shield of radius $a$ and thickness $t$ is~\cite{Lee2008}
\begin{equation}
    B_{{\rm n, curr}} = \mu_0 \sqrt{0.87\frac{k_{\rm B} T \,{\rm min}(t,\delta_{\rm skin})}{3\pi \rho a^2}},
    \label{Bnoise_shield}
\end{equation}
where $\delta_{\rm skin}$ is the skin depth and ${\rm min}(p,q)$ is the smallest of $p$ and $q$. Apart from a numerical factor, this is the same as the result for the slab given in (\ref{Bnoise_slab}) in the limit $t\ll z$. This result is for the noise component along the axis of the shield, which is our $y$-axis, but we suppose the noise in the $z$ direction will be similar since the fluctuating currents can be in any direction (this is also confirmed by our numerical studies in the next section). Lee and Romalis also give the ratio of the magnetization- and current-induced noises,
\begin{equation}
    \frac{B_{\rm n,magn}}{B_{\rm n,curr}} = \sqrt{\frac{3\mu''}{2\mu'}}\frac{\delta_{\rm skin}}{t},
    \label{Bnoise_ratio}
\end{equation}
where $\mu'$ and $\mu''$ are the real and imaginary parts of the permeability of the shielding material. 

Figure \ref{fig:ShieldNoiseAnalytical} shows the noise spectrum due to the inner magnetic shield at room temperature, calculated using Eqs.~(\ref{Bnoise_shield}) and (\ref{Bnoise_ratio}), assuming the two noise sources add in quadrature. The dimensions of the inner shield are $a=200$~mm and $t=1$~mm, and we have taken $\rho = 5.9 \times 10^{-7}~\Omega$~m, $\mu'=5\times 10^4 \mu_0$ and $\mu''=10^3 \mu_0$, close to the values reported in Ref.~\cite{Kornack2007} for a mu-metal shield. At frequencies below 0.1~Hz, magnetization noise dominates, scaling as $f^{-1/2}$. Between 0.1 and 3~Hz, Johnson noise dominates and there is no suppression from the skin effect. This is the most important region of the noise spectrum because the Fourier components of the electric field switching waveform lie in this part. Here, the noise is $5$~fT/$\sqrt{{\rm Hz}}$, almost independent of frequency and small enough that it is not a limiting factor in the EDM measurement. At $f>3$~Hz the noise is suppressed by the skin effect and falls off as $f^{-1/4}$. Note that the skin depth scales as $\mu^{-1/2}$ so this suppression is much more important for materials of high permeability.

\subsection{Magnetic noise from conductors and shields --- numerical calculations}

\begin{figure}[tb]
    \centering
    \begin{subfigure}{0.4\textwidth}
    \centering
        \includegraphics[width=0.8\textwidth]{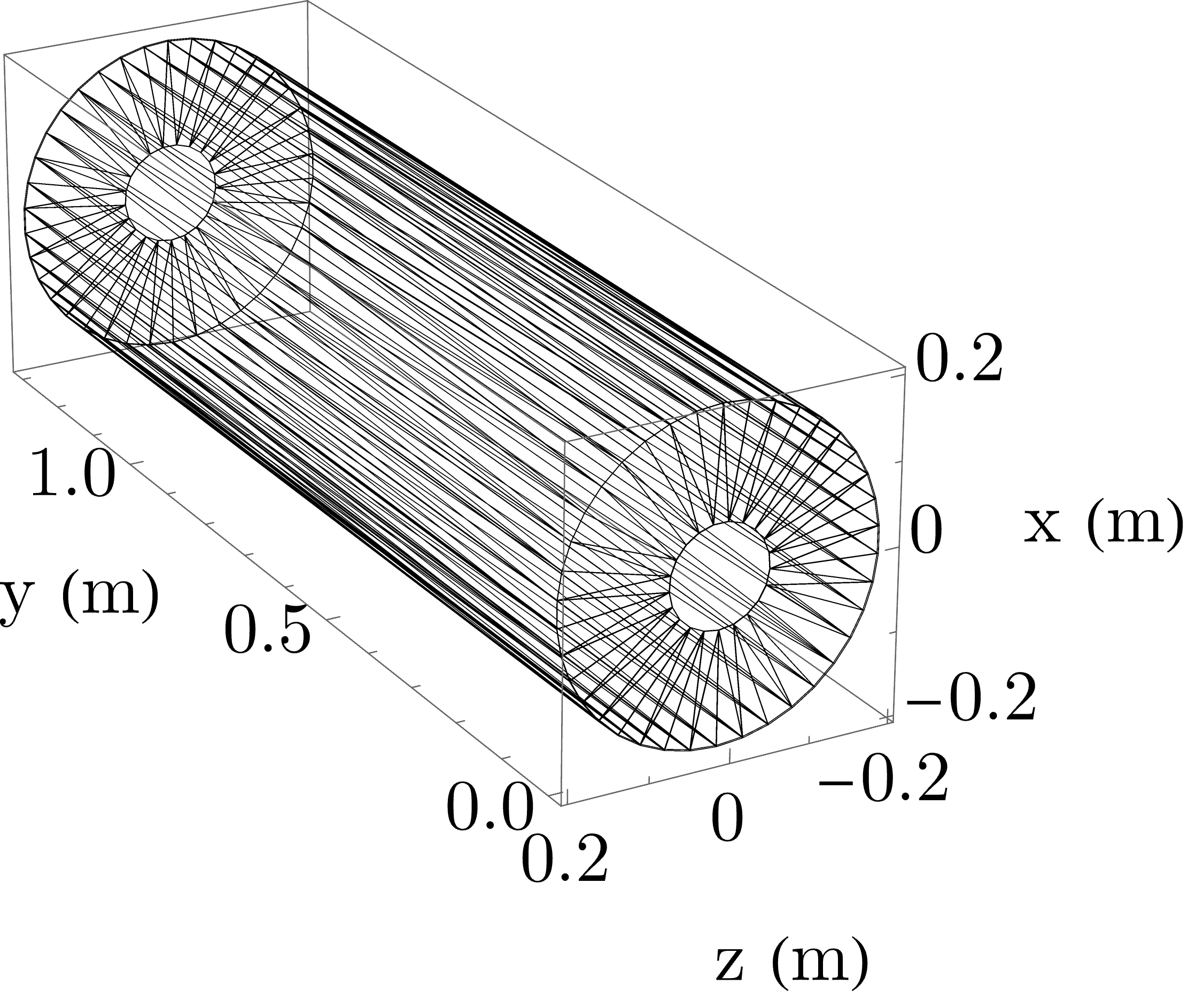}
        \caption{}
        \label{fig:ShieldMesh}
    \end{subfigure}\hfill
    \begin{subfigure}{0.6\textwidth}
    \centering
        \includegraphics[width=0.8\textwidth]{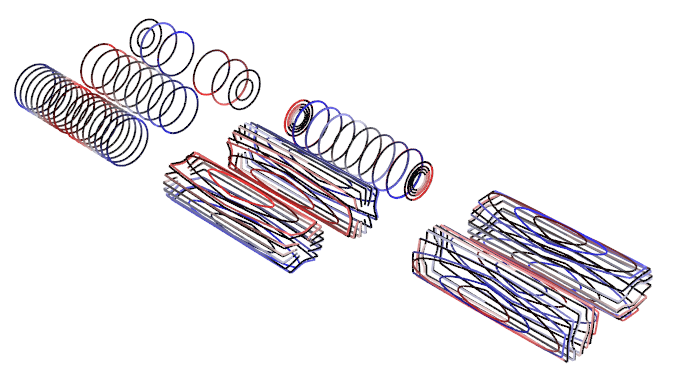}
        \caption{}
        \label{fig:ShieldCurrentModes}
    \end{subfigure}
    \begin{subfigure}{0.4\textwidth}
    \centering
        \includegraphics[width=0.8\textwidth]{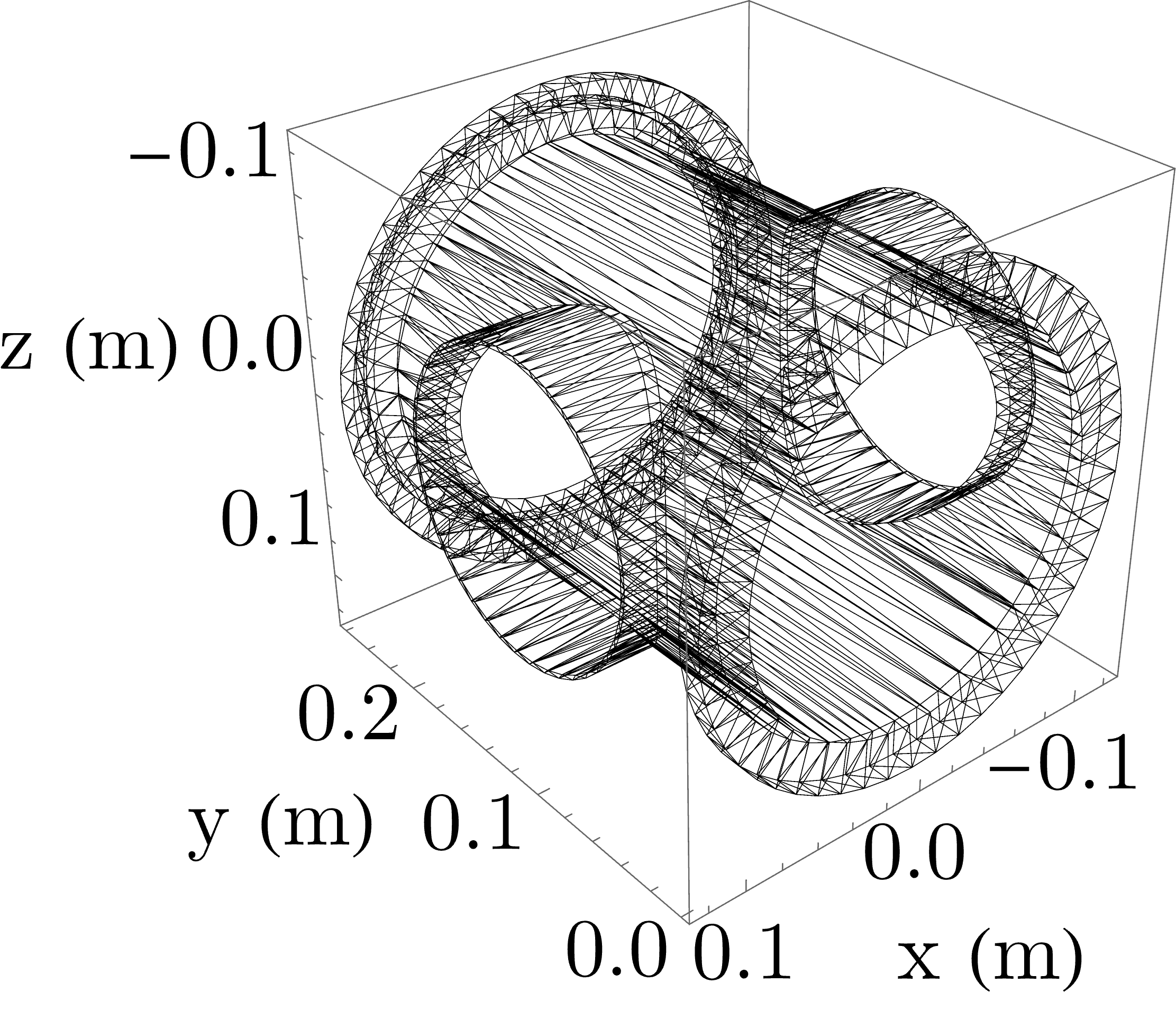}
        \caption{}
        \label{fig:TiChamberMesh}
    \end{subfigure}\hfill
    \begin{subfigure}{0.6\textwidth}
    \centering
        \includegraphics[width=0.8\textwidth]{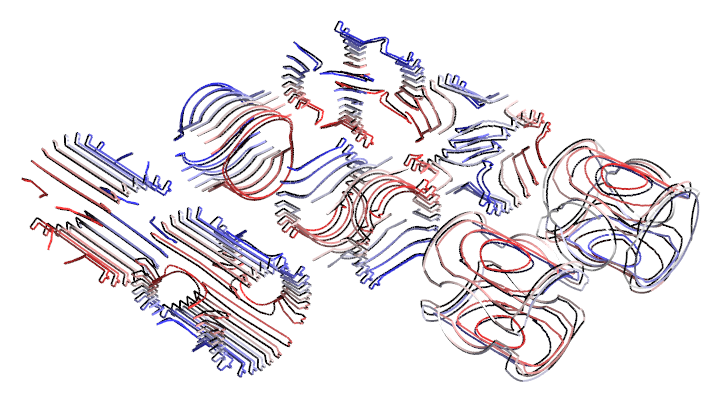}
        \caption{}
        \label{fig:TiChamberCurrentModes}
    \end{subfigure}
    \caption[]{Geometries used for numerical calculations of magnetic Johnson noise. (a, c) show the meshes used to model the inner layer of the mu-metal shield and the titanium chamber. (b, d) show some of the corresponding thermal current modes calculated for these geometries. Blue and red contours are the surface currents flowing in opposite directions.}
    \label{fig:JohnsonNoiseMeshes}
\end{figure}

In addition to the analytical estimates above, we have calculated the magnetic Johnson noise in the interaction region numerically, using a software package developed for this purpose~\cite{Iivanainen2021}.  This package computes the linearly independent modes of surface-current density for the given geometry, calculates the magnetic field associated with each of these patterns of surface current, and associates a mean thermal energy of $\frac{1}{2} k_{\rm B} T$ with each~\cite{Gillespie1996}. The magnetic permeability of the material is not included, but this makes only a small modification to the noise, as seen above and as noted in \cite{Lee2008}. Figure~\ref{fig:JohnsonNoiseMeshes} shows the geometries used to model the inner shield and the titanium vacuum chambers located at either end of the interaction region, along with the 8 lowest-order surface current modes. In calculating the noise, we use a few hundred modes which ensures that the results are converged to within 1\%. Once again, we focus on low frequencies where the Johnson noise is independent of frequency. 

\begin{figure}[tb]
    \centering
    \includegraphics[width=0.75\textwidth]{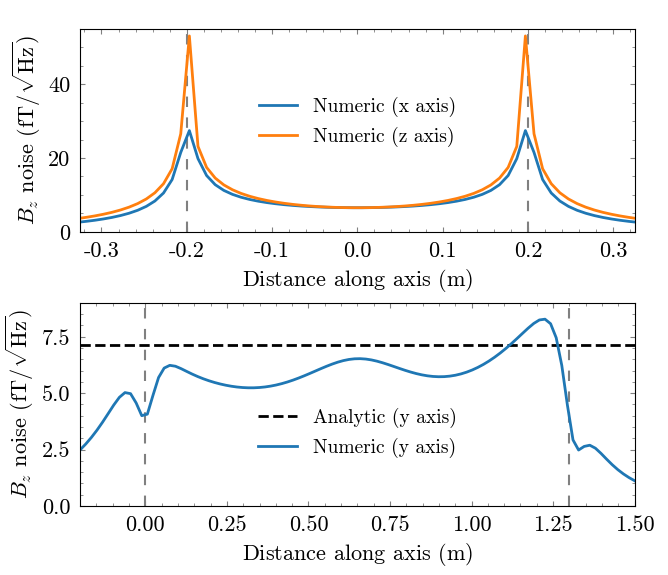}   
        \caption[]{Thermal $B_z$ noise along the three axes of the inner shield. The solid lines are calculated numerically and the dashed line is the analytical result for a conducting cylinder~\cite{Lee2008}. The shield has a length of 1.3~m and a diameter of 0.4~m. Top: Along $x$ and $z$. Bottom: Along $y$. Dashed vertical lines show the position of the mu-metal material.}
    \label{fig:ShieldDCNoise} 
\end{figure}

Figure~\ref{fig:ShieldDCNoise} shows the Johnson noise, $B_{{\rm n},z}$, produced by the inner shield. The noise is plotted along three orthogonal axes that pass through the centre point of the shield. Along the $y$-axis of the shield, the noise calculated numerically is always close to the value calculated analytically using the formula for a conducting cylinder~\cite{Lee2008}, which is 7.2~fT/$\sqrt{{\rm Hz}}$. The noise along the $x$ and $z$-axes also has this value near the centre, rising to about 30~fT/$\sqrt{{\rm Hz}}$ at the radius of the shield.

\begin{figure}[tb]
    \centering
    \includegraphics[width=0.75\textwidth]{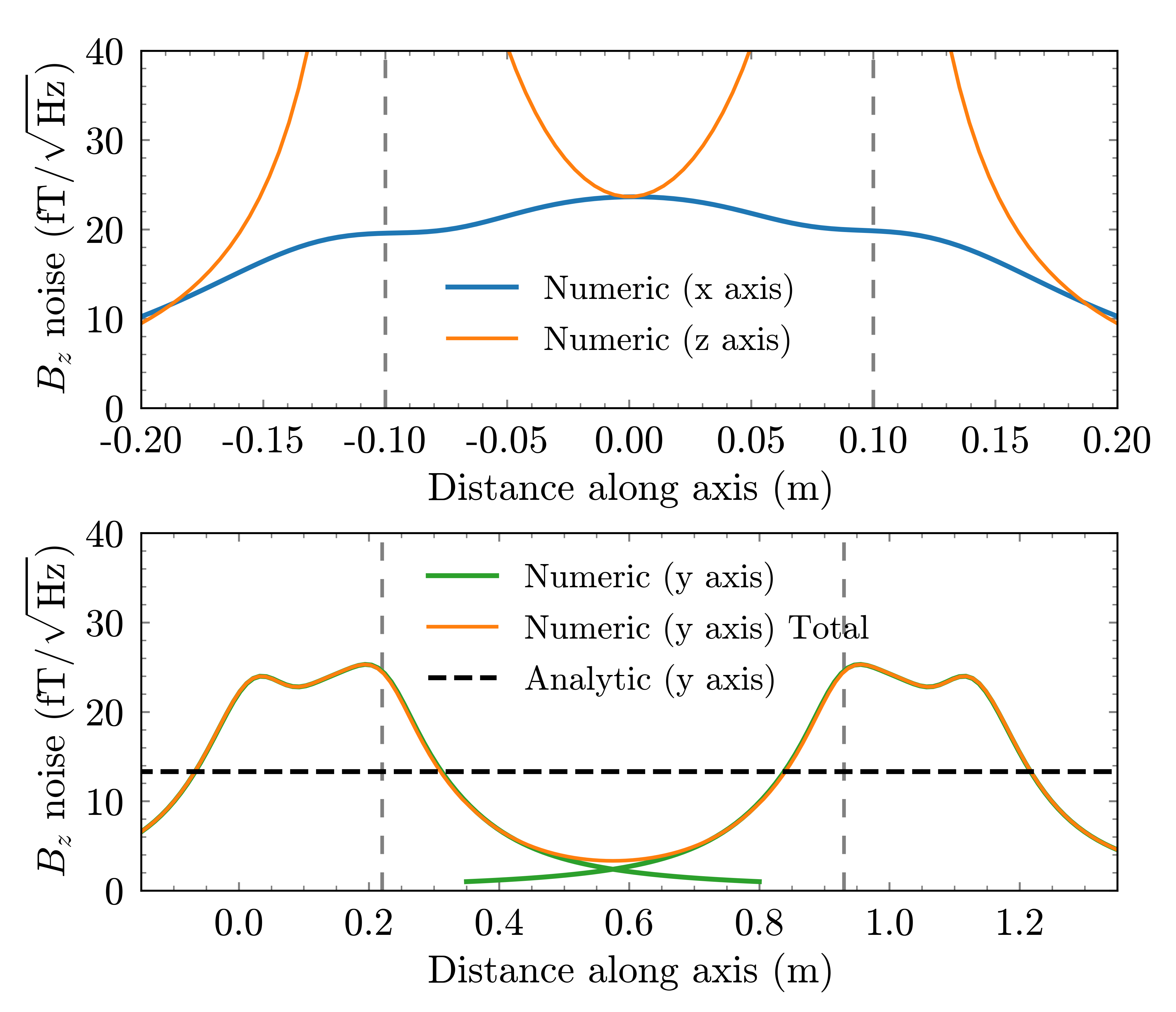}
    \caption[]{Thermal $B_z$ noise along the three axes of the titanium chambers. The solid lines are calculated numerically and the dashed line is the analytical result for a conducting cylinder~\cite{Lee2008} of the same radius as the chamber. Top: Noise along axes parallel to $x$ and $z$ that pass through the centre of a chamber. Vertical dashed lines indicate the walls of the chamber. Bottom: Noise along the $y$-axis due to individual chambers (green) and the two chambers together (orange). Vertical dashed lines show where the titanium chambers meet the glass tube.}
    \label{fig:TiChamberDCNoise}
\end{figure}

Figure~\ref{fig:TiChamberDCNoise} shows the Johnson noise produced by the titanium vacuum chambers at the two ends of the glass vacuum tube. The bottom plot shows the noise along the $y$-axis. It is very small at the centre of the interaction region, increasing to about 25~fT/$\sqrt{\text{Hz}}$ inside the chambers. The average value of the noise between the two STIRAP regions is 9.7~fT/$\sqrt{\text{Hz}}$. Increasing the length of the interaction region will further reduce this value. The top plot in figure~\ref{fig:TiChamberDCNoise} shows the noise along the two transverse axes passing through the centre of the titanium chamber. The noise increases towards the walls of the chamber, but there is little variation over the central 1~cm region of the chamber occupied by the molecules.

Using this numerical method, we have calculated the Johnson noise due to all parts of the apparatus inside the inner shield. The results are presented in table~\ref{tab:JohnsonNoiseTotals}. Here, we also show the EDM uncertainty per shot of the experiment due to the magnetic noise, which is $\sigma_{d_e} = \mu_{\rm B} B_{{\rm n},z}/(\sqrt{2\tau}E_{\rm eff})$. For this calculation, we have assumed $E_{\rm eff} = 18$~GV/cm and $\tau=20$~ms. This simple result is appropriate because the noise is independent of frequency so the modulation of the electric field direction is not relevant. We see from the table that the inner shield and titanium vacuum chambers contribute most of the noise, and that the EDM uncertainty due to Johnson noise is 5 times below the anticipated shot noise limit of the experiment.

\begin{table}[tb]
    \centering
    \begin{tabular} {ccc} \toprule
       Component &  $B_{{\rm n},z}$   & $\sigma_{d_e}$  per shot\\
       & (\SI{}{\femto\tesla\per\sqrt{\hertz}}) & ($10^{-29}$\EDMunits) \\
       \hline
        Field plates & 2.8 & $4.5 $ \\
        Ti joining plates & 2.9 & $4.7$ \\
        Inner shield & 6.1 & $10 $ \\ 
        Glass tube & 0 & 0 \\ 
        Ti chambers & 9.7 & $16$ \\  
        Transmission line & 0.40 & $0.64$ \\ 
        $B$-field coils & 0.12 & $0.19$ \\
        \midrule
        \textbf{Total} & 12.2 & $20.0$ \\ \bottomrule
    \end{tabular}
    \caption[]{Magnetic Johnson noise contributed by various components of the apparatus, averaged along the path taken by the molecules between STIRAP regions, with the associated EDM uncertainty per shot of the experiment. To calculate $\sigma_{d_e}$ we have assumed  $E_{\rm eff} = 18$~GV/cm and $\tau=20$~ms. The total noise is the quadrature sum of each contribution.}
    \label{tab:JohnsonNoiseTotals}
\end{table}

\subsection{Magnetic shielding measurements}

The laboratory magnetic noise is about 10~nT/$\sqrt{\rm Hz}$ at 0.1~Hz, falling to approximately 1~nT/$\sqrt{\rm Hz}$ at 1 Hz.  To ensure that this noise is not limiting, the shielding factor should be at least $10^5$ and preferably $10^6$. To measure the shielding factor of the four-layer shields, we pass an ac current of frequency $f$ through Helmholtz coils outside the shields. There is one coil pair for each axis. We measure the current and calculate the applied field, $\vec{B}_{{\rm ext}}$, at the position of the internal magnetometers. The array of magnetometers placed around the glass vacuum tube record the attenuated magnetic field, $\vec{B}_{{\rm int}}$. This determines the shielding factors, $S_{ij} = B_{{\rm ext},i}/B_{{\rm int},j}$, at various frequencies and directions, with $i,j \in \{x,y,z\}$.

\begin{figure}[tb]
    \centering
    \includegraphics[width=0.8\textwidth]{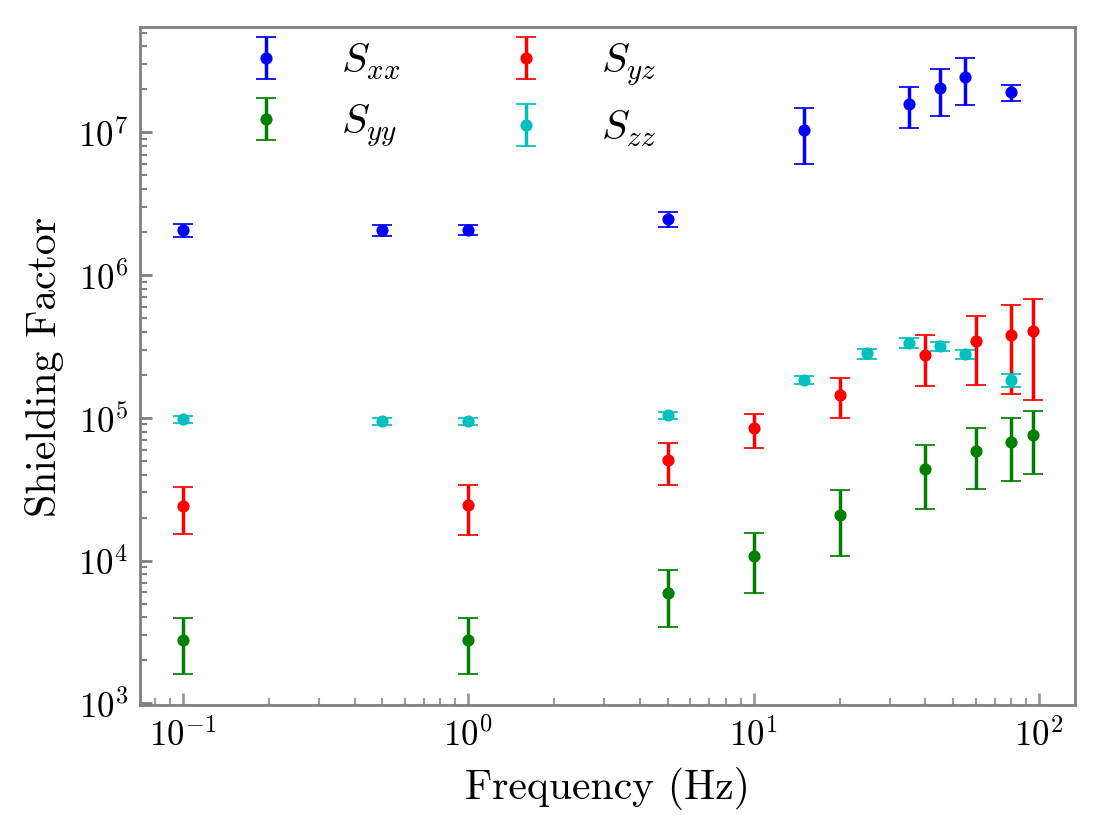}
    \caption[]{Measured shielding factors versus frequency. Points are the mean values determined from the number of internal magnetometers that measure the relevant axis and error bars are their standard deviations.}
    \label{fig:SFvsFreq}
\end{figure}
    
Figure \ref{fig:SFvsFreq} shows the results for several components of the shielding factor. We plot the mean and standard deviations of $S_{ij}(f)$ determined from the magnetometers. In all cases, the shielding factors are constant below a few Hz, then increase with frequency up to 50~Hz. Some components start to drop again at $f>50$~Hz. We focus on the values at low frequency. Here, we find $S_{xx} = 2.0 \times 10^6$, a very large value and more than adequate for our needs. However, in the $z$-direction, which is the most critical direction, we measure $S_{zz} = 1.0 \times 10^5$, which is only just adequate. The reason for this difference is explored in Sec.~\ref{Sec:seams}. The shielding is much worse again in the $y$-direction, with $S_{yy}=3 \times 10^3$ at low frequencies. This is not surprising given the geometry of the shields and is not directly of concern since the molecules are not sensitive to magnetic fields in this direction. More concerning is $S_{yz}$ which is about 4 times smaller than $S_{zz}$. This implies that the internal $B_z$ is actually dominated by the external magnetic field in the $y$ direction, and that magnetic noise may enter in the $y$ direction and be turned by the shields into noise in the $z$ direction. In section \ref{sec:B_noise_measure} we confirm that this is indeed the case.

\subsection{Seams}\label{Sec:seams}

\begin{figure}[tb]
    \begin{subfigure}{0.5\textwidth}
        \includegraphics[width=\textwidth]{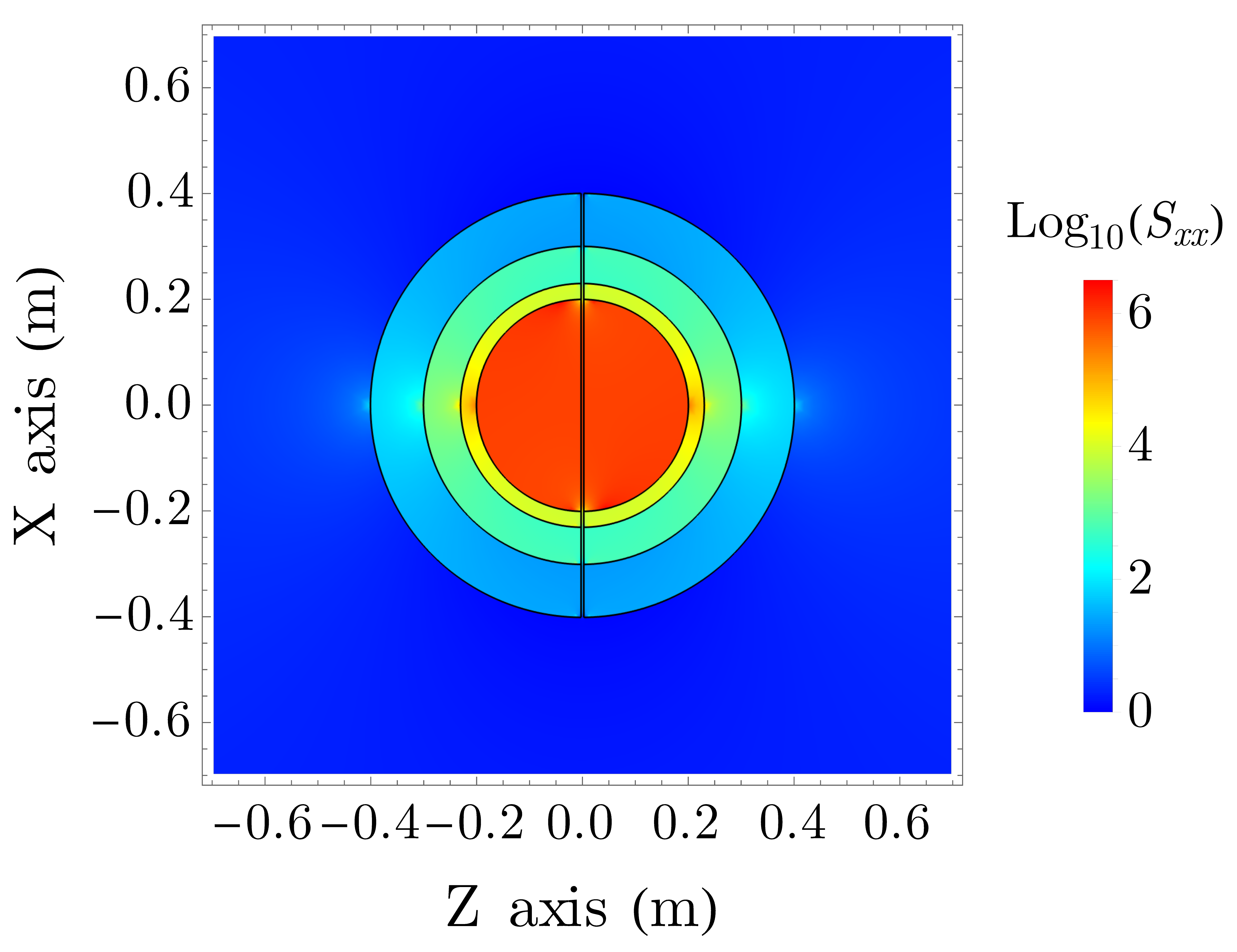}
        \caption{}
        \label{fig:ComsolModelSeam_Vertical}
    \end{subfigure}
    \begin{subfigure}{0.5\textwidth}
        \includegraphics[width=\textwidth]{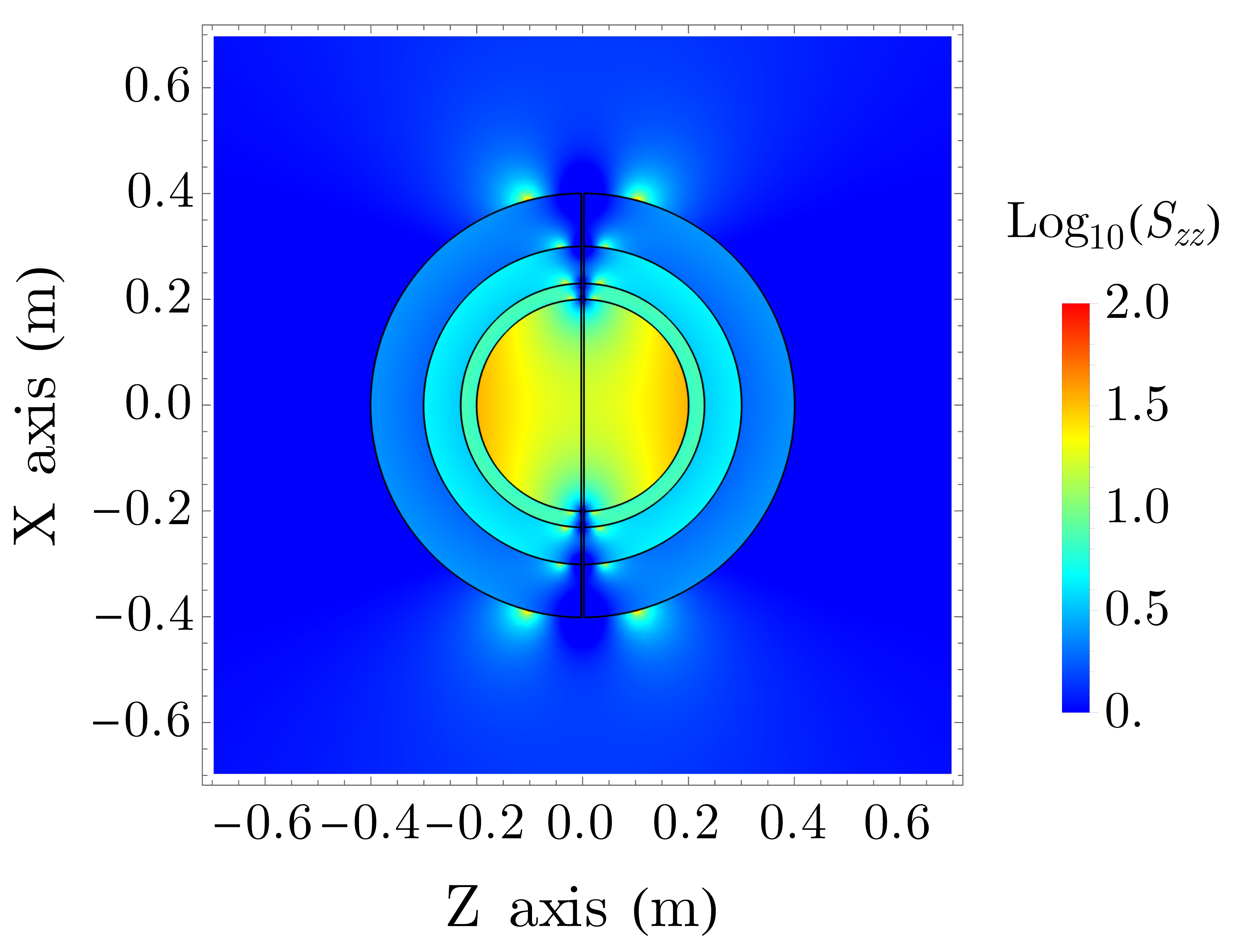}
        \caption{}
        \label{fig:ComsolModelSeam_Horizontal}
    \end{subfigure}
    \caption[]{Shielding factors determined from finite element modelling for a 4 layer shield with each cylinder split in half along the $x$ direction and a 5~mm gap between these halves. (a) Field applied in the $x$ direction. (b) Field applied in the $z$ direction. The colour scale shows a base 10 logarithm of the shielding factor.}
    \label{fig:ComsolModelSeams}
\end{figure}

The large difference in shielding performance for the two radial axes is related to the way the shield sections join together. The shields are half cylinders, split along their lengths and joined together with mu-metal bands parallel to the $y$-axis and at $z=0$ (i.e. at the top and bottom of each shield in figure \ref{fig:CADInteractionRegion}). To help understand the effect of these seams in the shielding material, we built a finite element model of the shields. When the shields are complete cylinders without seams, the model predicts a shielding factor of $10^7$. To exaggerate the effect of the seams, we then modelled the case where there is a 5~mm gap along the vertical ($x$) axis, and no mu-metal bands to cover the gap. Figure~\ref{fig:ComsolModelSeams} shows the calculated shielding factors for fields applied along $x$ and along $z$. We find $S_{xx} \approx 6\times 10^6$ and $S_{zz} \approx 2 \times 10^1$. Note carefully that the gap is along $x$ but the shielding factor is poor for fields directed along $z$. The shield is supposed to re-direct the magnetic field through the material instead of through the enclosed volume. When the field is applied along $x$ there is an un-interrupted path for the field lines to flow into the material, around the shield, and out the other side, so the effect of the gap is not too severe. When the field is applied along $z$ the path is interrupted by the gap, and the shielding factor is reduced enormously. We have also modelled the effect of adding mu-metal bands covering the gaps, but placed 1~mm away from the shield surface so that the material is still not quite continuous. This restores $S_{xx}$ to about $10^7$, the same value as found for complete cylinders, and increases $S_{zz}$ to about $10^5$. Furthermore, the modelling shows that a wider band is more effective at increasing $S_{zz}$ than a narrow band. Finally, we modelled the effects of the holes in the shields that are used to provide access to the beamline vacuum pipe (120~mm diameter), the STIRAP ports (50~mm diameter) and the high-voltage feedthroughs (50~mm diameter). The effect of these holes on $S_{zz}$ is small. A model that includes the holes but not the seams predicts $S_{zz} \sim 10^7$.

The gap between our shields is small, certainly less than 1~mm, and the bands that cover the seam are tightly connected to the main body of the shields with screws. Visually, the material appears continuous, but our measurements show that the shielding factor is 20 times worse in the $z$-direction and the modelling suggests that the seams are the cause. The modelling also shows that wider bands are more effective, so we hypothesize that the bands are currently too narrow and don't provide enough contact area for the flux to be routed efficiently through them. These results provide a cautionary tale for budding designers of magnetic shields -- while large holes can often be tolerated, invisible gaps that break the magnetic circuit are potentially ruinous. We note that seamless cylindrical shields are available, but these present many engineering challenges. The entire beamline would have to be inserted into the shields from one end which requires more space and a new concept for supporting the beamline. The shield diameter would have to be large enough to fully enclose high voltage feedthroughs, vacuum pumps and other large items attached to the vacuum chamber; that may be unfeasible in which case the vacuum would need to be broken and components removed in order to remove the shields.

\subsection{Magnetic noise measurements}\label{sec:B_noise_measure}

\begin{figure}[tb]
    \centering
    \includegraphics[width=\textwidth]{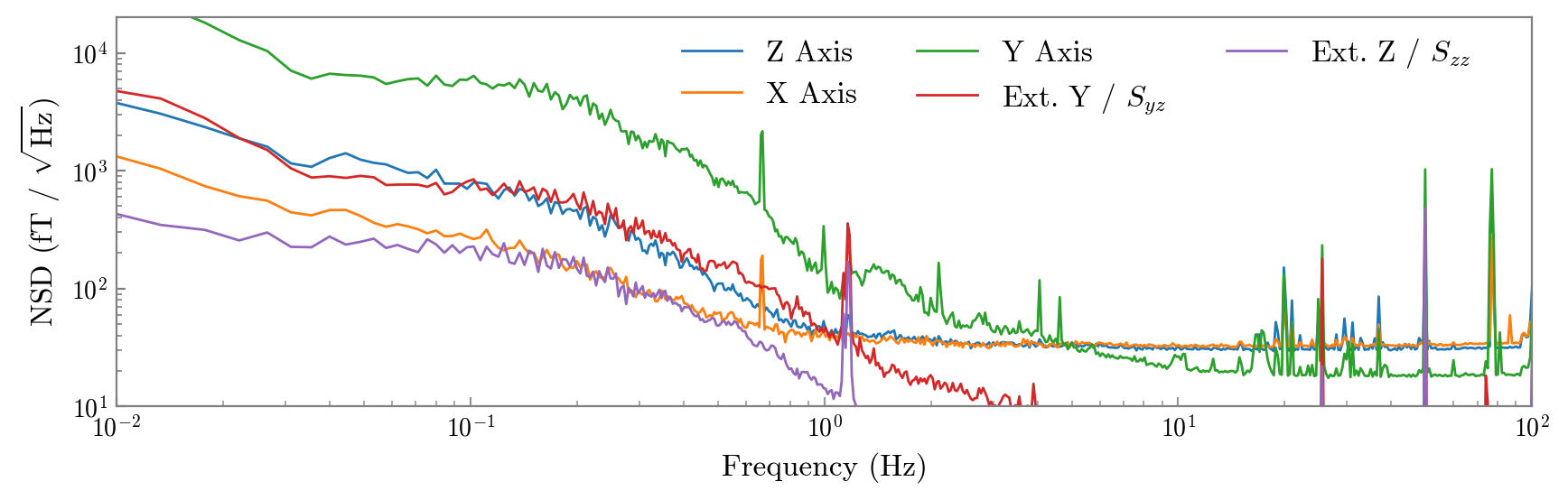}
    \caption{Noise spectral densities of the magnetic field measured inside the magnetic shield for all three axes. Included is the measured external noise along $z$ divided by the measured shielding factor $S_{zz}$, and the external noise along $y$ divided by $S_{yz}$.}
    \label{fig:NoiseSpectraUEDMZoomed}
\end{figure}

Figure \ref{fig:NoiseSpectraUEDMZoomed} shows the spectral density of the magnetic noise measured by the magnetometers inside the shields along all three axes. Data from each magnetometer are recorded for 30 minutes at a sample rate of 250~Hz and the noise spectral density of each is determined using the method described in Ref.~\cite{Welch1967}. For each axis, we average together the results from all magnetometers sensitive to that axis. 

Since $\phi$ is only sensitive to $B_z$, we are mainly concerned with the noise in the $z$-direction, $B_{{\rm n},z}$, which is shown by the blue line in the figure. It reaches a floor of 30~fT/$\sqrt{\rm{Hz}}$ at frequencies $f>3$~Hz. We attribute this to the noise of the magnetometers themselves; it is a little higher than the manufacturer's specification, but we find that the performance differs between sensors and that some do reach the specified value of 15~fT/$\sqrt{\rm{Hz}}$ in this frequency range. These differences between sensors are also responsible for the minor differences in the high-frequency noise floor measured in different directions.  At lower frequencies, where the noise floor of the magnetometers is not known, the total noise we measure starts to rise. It reaches 40~fT/$\sqrt{\rm{Hz}}$ at 1~Hz, 100~fT/$\sqrt{\rm{Hz}}$ at 0.4~Hz and 500~fT/$\sqrt{\rm{Hz}}$ at 0.1~Hz. In section \ref{Sec:NoiseImpact} we found that magnetic noise should be below 100~fT/$\sqrt{\rm{Hz}}$ at the switching frequency of the electric field, so that this noise does not limit the sensitivity. Thus, we see that the noise is sufficiently small provided the electric field is switched at 0.4~Hz or higher. To quantify this more carefully, we determine $\phi_E$ (the uncertainty in the phase correlated with the modulation of $E$) using (\ref{eq:sigma_phiE_magnetic}), together with the measured noise spectrum (figure~\ref{fig:NoiseSpectraUEDMZoomed}) and the $E$ switching waveform used in the experiment whose fastest frequency is 1~Hz. Taking $\tau=20$~ms, this analysis gives $\sigma_{\phi_E}= 3.5 \times 10^{-4}$~rad per shot, 3 times below the best possible sensitivity when there are $10^6$ molecules per shot.

The purple line in figure~\ref{fig:NoiseSpectraUEDMZoomed} shows the noise in $B_z$ measured outside the shields, divided by the shielding factor $S_{zz}$. In the low frequency region of interest, the result is about 5 times smaller than the internal $B_z$ noise we actually measure. This implies that $B_{{\rm n},z}$ is not limited by the shielding in the $z$-direction. The red line shows the external noise in $B_y$ divided by $S_{yz}$, and this line lies very close to the noise in $B_z$ at all frequencies below 1~Hz. This shows that $B_{{\rm n},z}$ is limited by the shielding factor in the $y$-direction. Rather than passing directly through the shields in the $z$-direction, the noise enters in the $y$ direction and is then turned by the shields into noise in the $z$ direction.

The orange line in figure~\ref{fig:NoiseSpectraUEDMZoomed} shows the noise in $B_x$. This is almost identical to the noise in $B_z$ for frequencies above 0.7~Hz, presumably because the measurements are limited by the noise floor of the magnetometers at these frequencies. At lower frequencies, the noise in $B_x$ is lower than the noise in $B_z$, but only by a factor of 3, even though the shielding factor is 20 times better in the $x$-direction. Moreover, the internal noise in $B_x$ is about 100 times larger than the external noise in $B_x$ divided by $S_{xx}$, again suggesting that the noise enters indirectly, rather than through the radial body of the shields. The green line in the figure shows the noise in $B_y$. Above 10~Hz it is 20~fT/$\sqrt{\rm{Hz}}$, limited by the noise floor of the sensors. Below 1~Hz, it is 3-10 times larger than the noise in $B_z$. This is unsurprising given the low shielding factor measured in the $y$-direction (see figure \ref{fig:SFvsFreq}).

In summary, we find that the low frequency magnetic noise in the most important direction ($z$) is dominated by external noise that enters the shields in the $y$ direction where the shielding is poor, and is then turned into noise in the $z$ direction by the shields. It is likely that this noise could be reduced substantially by reducing the axial holes in the shields (with a corresponding reduction in the diameter of the vacuum tube), improving the connection between axial sections of the shields, and/or implementing active reduction of the external magnetic field in the axial direction.

\subsection{Residual magnetic field and field gradients}

The phase $\phi$ given by (\ref{eq:phi}) is proportional to the magnetic field integrated along the molecule trajectory. The background magnetic field, and its gradients, need to be controlled carefully. After degaussing the shield, the internal magnetometers, placed 115~mm from the axis, read an average field of about 2~nT. The magnetic field measured on axis by the molecules is similar. We use the internal coils to apply an offset field that cancels this residual background. This cancellation works well. For example, during a data run lasting 2.5 hours the background magnetic field had a mean value of $2.5 \pm 2.5$~pT.

Magnetic field gradients produce different phases for different trajectories, resulting in a spread of phases that reduces sensitivity. Consider a molecule whose trajectory is $\{x(y),z(y)\} = \{x_0+\frac{v_x}{v_y}y-g\frac{y^2}{2v_y^2}, z_0+\frac{v_z}{v_y}y\}$, where $\{x_0,z_0\}$ is the initial transverse displacement, $\{v_x,v_z\}$ is the transverse velocity, and $g$ is the acceleration due to gravity. Using a Taylor expansion of $B(x,y,z)$ about the origin, and integrating over the trajectory, the phase accumulated by this molecule is
\begin{equation}
    \phi = \frac{\mu_{\rm B}L}{\hbar v_y}\left[\left(\!B_{z,0}\!+\!\frac{1}{2}\frac{\partial B_z}{\partial y}L\!\right)\!+\!\left(\!\frac{\partial B_z}{\partial x}x_0\!+\!\frac{\partial B_z}{\partial z}z_0\!\right)\!+\!\left(\!\frac{\partial B_z}{\partial x}v_x\!+\!\frac{\partial B_z}{\partial z}v_z\!\right)\frac{L}{2v_y}\!-\!\left(\!\frac{1}{6v_y^2}\frac{\partial B_z}{\partial x}g L^2\!\right)\right].
    \label{eq:B_gradient_phase}
\end{equation}
The first bracket contributes to the phase spread through the spread in forward velocities ($\Delta v_y$). The second contributes through the spread of initial positions ($\Delta x_0,\Delta z_0$), the third through the spread of transverse velocities ($\Delta v_x,  \Delta v_z)$, and the fourth through the variation in the gravitational drop at various forward velocities.

The bias field $B_{z,0}$ is chosen to give a mean phase $\phi_0=\pi/4$. We will see shortly that we prefer $B_{z,0}$ to dominate in the first bracket, so we can write this condition in the approximate form $\phi_0 = \mu_{\rm B} B_{z,0}L/(\hbar \langle v_y\rangle)=\pi/4$. We can set limits on the magnetic field gradients by requiring $\Delta \phi/\phi_0 < 1/\zeta$ where $\zeta$ is a number that we will choose in a moment. This results in the following limits:
\begin{align}
    &\left|\frac{\partial B_z}{\partial x}\right| < \left\{\frac{1}{\zeta}\frac{B_{z,0}}{\Delta x_0},\frac{2}{\zeta}\frac{B_{z,0}v_y}{L\Delta v_x}, \frac{2}{\zeta}\frac{B_{z,0}v_y^3}{g L^2 \Delta v_y}\right\},\label{eq:gradientLimX}\\
    &\left|\frac{\partial B_z}{\partial z} \right|< \left\{\frac{1}{\zeta}\frac{B_{z,0}}{\Delta z_0},\frac{2}{\zeta}\frac{B_{z,0}v_y}{L\Delta v_z} \right\}. \label{eq:gradientLimZ}
\end{align}
To choose an appropriate $\zeta$, we note that the first term in (\ref{eq:B_gradient_phase}) is independent of the gradients and results in $|\Delta \phi/\phi_0 | = \Delta v_y/v_y$. The magnetic gradients will not have much impact on the sensitivity if they generate phase spreads smaller than this, so it is natural to choose $1/\zeta=\Delta v_y/v_y$. For the beam currently used in the experiment, this value is $1/\zeta\approx 1/4$.

The constraint on $\frac{\partial B_z}{\partial y} $ has a different source. Some systematic errors in the state preparation and analysis scale with the applied magnetic field~\cite{Ho2023c}, which in this case means the magnetic field in the STIRAP regions. The field in these regions should not be much larger than the uniform field $B_{z,0}$, leading to the constraint
\begin{equation}
\left| \frac{\partial B_z}{\partial y} \right|< \frac{2 B_{z,0}}{L}.
\label{eq:gradientLimY}
\end{equation}

\begin{table}[tb]
    \centering
    \begin{tabular} {ccccc} \toprule
       Gradient &  Position spread   & Velocity spread & Gravity & Offset\\
       \hline
        \multirow{2}{2em}{$\left|\frac{\partial B_z}{\partial x}\right|$} & $1.4 \times 10^{-7}$  & $3.4 \times 10^{-6}$ & $1.3 \times 10^{-5}$ & -- \\
        & $7.4 \times 10^{-9}$  & $8.9 \times 10^{-9}$ & $1.8 \times 10^{-9}$ & -- \\
        \hline
        \multirow{2}{2em}{$\left|\frac{\partial B_z}{\partial y}\right|$} & -- & -- & -- & $4.5 \times 10^{-9}$ \\
         & -- & -- & -- & $6.0 \times 10^{-11}$ \\
         \hline
        \multirow{2}{2em}{$\left|\frac{\partial B_z}{\partial z}\right|$} & $1.4 \times 10^{-7}$  & $3.4 \times 10^{-6}$ & -- & -- \\
        & $7.4 \times 10^{-9}$  & $8.9 \times 10^{-9}$ & --  & --\\
         \bottomrule
    \end{tabular}
    \caption[]{Upper limits to magnetic field gradients, in T/m, due to spread of initial positions, spread of transverse velocities, spread of drops under gravity, and the requirement that the offset phase be dominated by $B_{z,0}$. Values are from Eqs.~(\ref{eq:gradientLimX}), (\ref{eq:gradientLimZ}), (\ref{eq:gradientLimY}),  with $\Delta x_0 = \Delta z_0 = 3$~mm, $\Delta v_x = \Delta v_z = 0.05$~m/s, $1/\zeta=\Delta v_y/v_y = 1/4$. In each case, the upper value is for an experiment with $\langle v_y \rangle =150$~m/s, $L=0.77$~m, while the lower is for $\langle v_y \rangle=30$~m/s, $L=3$~m.}
    \label{tab:GradientLimits}
\end{table}

Table \ref{tab:GradientLimits} gives the upper limits implied by Eqs.~(\ref{eq:gradientLimX}), (\ref{eq:gradientLimZ}) and (\ref{eq:gradientLimY}). We have taken $\Delta x_0 = \Delta z_0 = 3$~mm, and $\Delta v_x = \Delta v_z = 0.05$~m/s, which are typical values for our YbF beam when it is laser cooled to 50~$\mu$K. The upper values in the table are for the experiment as currently configured, which has $\langle v_y \rangle = 150$~m/s and $L=0.77$~m, requiring $B_{z,0}=1.7$~nT. Here, the limit on the transverse gradients is 140~nT/m, and comes from the spread of initial positions, while the limit on the longitudinal gradient is 4.5~nT/m. The lower values in the table are for a more ambitious experiment which has $\langle v_y \rangle = 30$~m/s and $L=3$~m, requiring $B_{z,0}=89$~pT. In this case, the limit on $\frac{\partial B_z}{\partial x}$ is 1.8~nT/m and comes from the spread in gravitational drops, the limit on $\frac{\partial B_z}{\partial z}$ is 7.4~nT/m and comes from the spread of initial positions, and the limit on $\frac{\partial B_z}{\partial y}$ is 60~pT/m.  

Each of the eight magnetometers measures the field along two axes. From this set of measurements we can estimate the gradients. We measure both $\partial B_z/\partial z$ and $\partial B_z/\partial x$ at the centre of the interaction region and at one end of the glass tube. Their mean values are $|\partial B_z/\partial z| = 14.7$~nT/m and $|\partial B_z/\partial x| = 7.4$~nT/m. For $\partial B_z/\partial y$ we use 4 gradients measured along the length of the glass tube, obtaining a mean value of $|\partial B_z/\partial y| = 1.1$~nT/m. The uncertainties on these measurements are roughly as large as the mean values themselves because the gradients are only sampled at a few points. Nevertheless, comparing to the upper set of values given in table \ref{tab:GradientLimits} it is clear that the gradients are small enough for the experiment as currently configured. For a more demanding experiment where $\langle v_y \rangle = 30$~m/s and $L=3$~m, the gradients in $x$ and $z$ are close to being adequate, but considerably more attention would be needed to control $\partial B_z/\partial y$ to the required level.

\subsection{Magnetic fields correlated with E-reversal}

\begin{figure}[tb]
    \centering
    \includegraphics[width=0.7\textwidth]{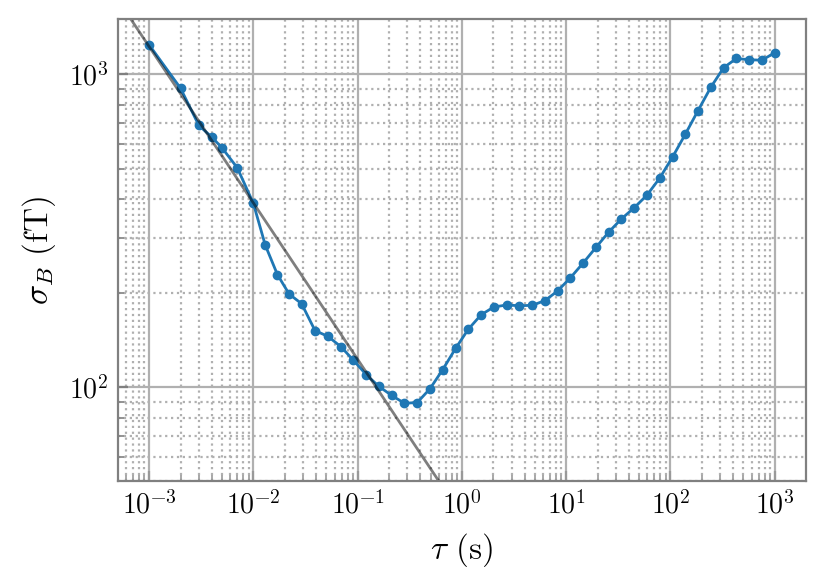}   
    \caption{Allan deviation of the magnetic field in the $z$-direction, $\sigma_B(\tau)$, measured inside the magnetic shield. We have averaged the results from all the magnetometers. The straight line is proportional to $1/\sqrt{\tau}$. }
    \label{fig:AllanDeviation_Zaxis}
\end{figure}

The EDM is determined from the change in phase correlated with the reversal of the electric field. If the magnetic field changes when the electric field is reversed, a systematic error arises. To combat this error, we measure this E-correlated magnetic field, $B_E$, using the array of magnetometers. Figure~\ref{fig:AllanDeviation_Zaxis} shows the sensitivity of these magnetometers in our setup, as represented by their Allan deviation~\cite{Allan1966}.
The Allan deviation, $\sigma_B(\tau)$, evaluated at averaging time $\tau$, is a measure for the instability between two consecutive $B$-field measurements a time $\tau$ apart, giving the root mean square value of the magnetic field fluctuations during the observation period $\tau$.
When the magnetic field measurements are dominated by white noise, the Allan deviation scales as $1/\sqrt{\tau}$.
Pink noise, or $1/f^{1/2}$ in the noise spectral density, appears as a flat trend in the Allan deviation, independent of $\tau$, whereas a random walk ($1/f$ in the noise spectral density) produces a trend proportional to $\sqrt{\tau}$. In figure~\ref{fig:AllanDeviation_Zaxis} we have averaged together the results from all the magnetometers. We see that the Allan deviation scales approximately as $1/\sqrt{\tau}$  (shown by the straight line) up to an averaging time of 300\,ms, where it reaches its minimum value of 90~fT. For longer $\tau$, random walk takes over, and $\sigma_B(\tau)$ starts to rise. The figure shows that, to measure $B_E$ as precisely as possible, it is best to choose an averaging time of about 300~ms.

\begin{figure}[tb]
    \centering
    \includegraphics[width=0.75\textwidth]{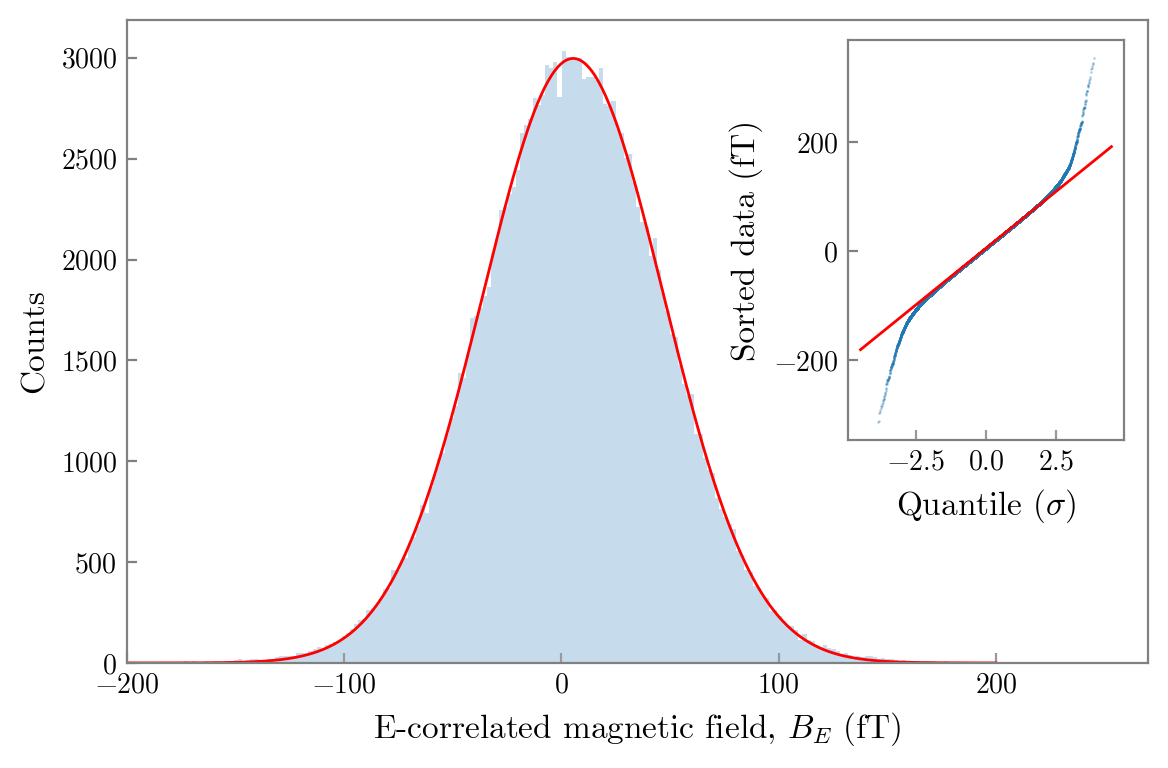}
    \caption{Histogram of the $E$-correlated magnetic field, $B_E$, obtained from one magnetometer. The uncertainty in this measurement is 0.11~fT after 104 hours. The inset shows the sorted values plotted against the quantiles of a Gaussian distribution. The red lines are fits to a Gaussian distribution.}
    \label{fig:ECorrelatedBHistogram}
\end{figure}

To characterize $B_E$, we acquire a series of magnetometer measurements with the $E$ field direction switched back and forth between $\pm 6.6$~kV/cm. The magnetometers are recorded simultaneously for 150~ms, with a dead time of 400~ms between readings where the electric field is switched and allowed to settle. As indicated in Figure~\ref{fig:AllanDeviation_Zaxis}, these timings balance the 100\,fT statistical sensitivity at 150\,ms with the systematic uncertainty due to random walks that start dominating after 300\,ms. 
The measurements were performed with a polarity switching pattern consisting of $+--+$ and $-++-$ units, reducing sensitivity to drifts of the background magnetic field. From each of these units, we determine a value of $B_E$. Figure~\ref{fig:ECorrelatedBHistogram} shows the distribution of these values for one of the magnetometers, from a measurement spanning 104 hours. The mean value measured by this magnetometer is $5.48\pm 0.11$~fT. All the magnetometers measure the field with a similar sensitivity to this one, although their central values differ. The uncertainty is determined from the standard deviation divided by the square root of the number of results. This is a good estimate because the measurements follow a normal distribution out to $\pm 2.5$ standard deviations, showing only a little excess noise in the wings of the distribution, as can be seen from the quantile plot shown in the inset of figure~\ref{fig:ECorrelatedBHistogram}. The ability to measure $B_E$ with a precision of 0.11~fT per magnetometer, in a relatively short time, is an important achievement. With this sensitivity, the systematic uncertainty due to $E$-correlated magnetic fields can be measured at the $10^{-31}~e$~cm level in a reasonable time frame. We are currently studying the source of the non-zero $B_E$ we measure. If the source is outside the shields, e.g. the high voltage switches, it will be straightforward to eliminate. More difficult are internal sources such as the residual leakage current, or electromagnetic interference at the magnetometers due to the changing electric field or the charging currents. Eliminating $B_E$ is a crucial task before EDM measurements can be made.

\section{Summary and conclusions}

Low energy tests of fundamental physics, such as measurements of the EDM, often require exceptionally precise control over electric and magnetic fields. In this paper, we have presented our efforts to create a region of high electric field and very low magnetic noise, suitable for making an EDM measurement using a beam of ultracold YbF molecules. 

Many of our design choices were driven by the need for very low magnetic noise, especially at the frequency of the electric field reversal, which is around 1~Hz in this experiment. By using ceramic electric field plates inside a glass vacuum chamber and a multi-layer mu-metal shield with an inner radius of 0.2~m, we estimate the Johnson noise to be 12.2~fT/$\sqrt{{\rm Hz}}$, remaining roughly independent of frequency at the low frequencies of relevance here. The main contributors to this noise are the titanium chambers at the ends of the glass tube, and the innermost layer of mu-metal. External magnetic noise enters the interaction region due to imperfect shielding. We measure 40~fT/$\sqrt{{\rm Hz}}$ at 1~Hz. This translates to a phase noise of 0.35~mrad per shot when the spin precession time is $\tau = 20$~ms, and 1.1~mrad per shot when $\tau = 200$~ms. Comparing this to the shot noise limit of $1/\sqrt{N}$ for $N$ particles, we see that the magnetic noise is not a limiting factor unless $N>8 \times 10^6$ per shot for $\tau = 20$~ms, or $N>8 \times 10^5$ per shot for $\tau = 200$~ms. Converting this magnetic noise to a limit on EDM sensitivity, we find that for $\tau =20$~ms and $E_{\rm eff} = 18$~GV/cm, the limit is $6.4 \times 10^{-28}~e$~cm in one shot. Assuming a typical data-taking rate of 10 shots per second and a 50\% duty cycle, this translates to $9.7\times 10^{-31}~e$~cm in one day.

While magnetic noise does not currently limit the sensitivity of the experiment, it may become significant in the future as new methods are introduced to increase $\tau$ and $N$. Then, improving the shielding of external magnetic noise may be necessary. The present cylindrical shield has a shielding factor of $2 \times 10^6$ in one radial direction ($x$), $1 \times 10^5$ in the other radial direction ($z$), and $3 \times 10^3$ in the axial direction ($y$). The large difference in shielding factors in the two radial directions is attributed to magnetic reluctance introduced by joints in the shields, inhibiting the shield's ability to redirect magnetic flux through the material even though the joints are covered by mu-metal bands. The poor shielding in the axial direction is partly due to the large axial holes in the shields where the vacuum tube passes through, and partly due to joints along the length of the shield. Fortunately, molecular EDM measurements are only sensitive to the magnetic field in the direction of the applied electric field, here the $z$-axis. However, we find that for all frequencies studied ($10^{-2}-10^2$~Hz), the noise in $B_z$ is significantly larger than the external magnetic noise divided by the shielding factor in the $z$-direction. Above 1~Hz, we attribute this to the noise floor of the magnetometers. Below 1~Hz, we attribute it to noise that enters in the $y$-direction, where the shielding is poor, and is partly re-directed by the shields into the $z$-direction. This problem may be exacerbated by the joints in the mu-metal. 

These conclusions contain important lessons for future shield designs. Potential future improvements include: (i) rotating the shields to align the largest shielding factor with the $z$ direction; (ii) using wider mu-metal bands to cover the joints and improving the way they are fastened; (iii) reducing the diameter of the vacuum tube where it passes through the shields so that the axial hole size can be reduced; (iv) active compensation of the external magnetic noise, especially the low frequency noise in the $y$-direction; (v) using seamless shields if the engineering challenges noted in section \ref{Sec:seams} can be overcome. Cooling of the shields is unlikely to offer any benefit because (a) reducing $T$ also reduces $\rho$, resulting in only a marginal reduction in Johnson noise and (b) the permeability of mu-metal, and hence its shielding factor, falls off very rapidly at temperatures below 270~K. Other types of nickel-iron alloys have much better performance at lower temperatures, see e.g. \cite{Zhou2024}.

We find that Al$_2$O$_3$ coated with a thin layer of TiN is an excellent material for electric field plates. These plates generate minimal magnetic Johnson noise, and due to their rigidity can be mounted with high parallelism and minimal bending. We have never observed electrical discharges at an electric field of 20~kV/cm. Patch potentials on the plates could potentially lead to a fluctuating geometric phase or a systematic error from a geometric phase correlating with the electric field direction. Examining a 30~mm x 30~mm area of the plates, we measured an RMS variation in the work function of 54~meV. This is too small to be of concern. 

A magnetic field correlated with the reversal of the electric field, $B_{E}$, is a systematic error. With $E_{\rm eff} = 18$~GV/cm, this error scales as $d_{e,{\rm false}}/B_E = 3.2 \times 10^{-30}~e$~cm/fT. Such an $E$-correlated magnetic field will be measured by the array of magnetometers that surround the glass vacuum tube. We have shown that a single magnetometer can measure $B_E$ with an uncertainty of $0.11$~fT in 104 hours of data taking, equivalent to a systematic EDM uncertainty of $3.6 \times 10^{-31}~e$~cm. To reach a systematic uncertainty below $10^{-31}~e$~cm, 57 days of magnetometry data are required. Alternatively, if spatial resolution is not important, the results from the 8 magnetometers can be averaged together so that this limit can be reached in about 7 days. Many potential sources of $B_E$ are external to the shields, so must be many orders of magnitude larger near their source, making them easy to detect using external magnetometers. The exception is leakage current between the electric field plates, which could produce a $B_E$ directly within the interaction region, making it harder to measure. At present, we measure a leakage current of approximately 620~pA when the applied field is 20~kV/cm, but the current path is unknown. If it takes a conspiratorial path, it could generate $B_E = 4$~fT at the molecules, leading to a systematic error of $1.3 \times 10^{-29}~e$~cm. This current should also produce a comparable $B_E$ at one or more of the internal magnetometers, so is unlikely to go undetected. Nevertheless, it would be beneficial to use co-magnetometry to determine the exact magnetic field seen by the molecules, for example by using two different states of the molecule~\cite{Fitch2020b}.

Our work offers a set of valuable techniques of benefit to other EDM experiments~\cite{Alarcon2022}, measurements of nuclear Schiff moments~\cite{Grasdijk2021, Arrowsmith-Kron2024Short} and magnetic quadrupole moments~\cite{Hutzler2020, Denis2020, Ho2023}, measurements of parity violation~\cite{Altuntas2018}, the determination of fundamental constants and their variation~\cite{Leung2022, Barontini2022Short}, and searches for new physics by atom interferometry~\cite{Abe2021Short, Badurina2020Short}.

The data supporting the findings in this paper are available at 10.5281/zenodo.15094949.

\section*{Acknowledgements}
We thank David Pitman for technical support. This work has been supported by the ``Table-top experiments for fundamental physics'' program, sponsored by the Gordon and Betty Moore Foundation (Grants 8864 \& GBMF12327, DOI 10.37807/GBMF12327), Simons Foundation, Alfred P. Sloan Foundation, and John Templeton Foundation Grants (G-2019-12505 \& G-2023-21035). The research has also been supported by UKRI under Grants EP/X030180/1, ST/V00428X/1 and ST/Y509978/1, and by the Royal Society (URF\textbackslash R1\textbackslash 180578).

\newcommand{\newblock}{}
\bibliographystyle{IEEEtran}
\bibliography{references}

\end{document}